\documentclass[aps,prd,nofootinbib,showpacs,preprintnumbers,longbibliography,amssymb,11pt]{revtex4-1} % not twocolumn
\usepackage[sort&compress]{natbib}

\usepackage{graphicx}% Include figure files
\usepackage{epsfig}
\usepackage{dcolumn}% Align table columns on decimal point
\usepackage{bm}% bold math
\usepackage{amsmath}
\usepackage{psfrag}%eps figure labels are placeholders for tex

\graphicspath{{./Figures/}}

\newcommand{\eqnref}[1]{Eq.~(\ref{eqn:#1})}

\newcommand{\secref}[1]{Sec.~\ref{sec:#1}}

\newcommand{\figref}[1]{Fig.~\ref{fig:#1}}
\newcommand{\figsref}[2]{Figs.~\ref{fig:#1} and \ref{fig:#2}}

\newcommand{\tableref}[1]{Table~\ref{table:#1}}
\newcommand{\tablesref}[2]{Tables~\ref{table:#1} and \ref{table:#2}}

\begin{document}

\preprint{FERMILAB-PUB-14-072-T}

\title{Anatomizing Exotic Production of the Higgs Boson}

\author{Felix Yu}
\email{felixyu@fnal.gov}
\affiliation{Theoretical Physics Department, Fermilab, 
Batavia, IL 60510, USA}

%\date{\today}

\begin{abstract}

We discuss exotic production modes of the Higgs boson and how their
phenomenology can be probed in current Higgs analyses.  We highlight
the importance of differential distributions in disentangling standard
production mechanisms from exotic modes.  We present two model
benchmarks for exotic Higgs production arising from
chargino-neutralino production and study their impact on the current
Higgs dataset.  As a corollary, we emphasize that current Higgs
coupling fits do not fully explore the space of new physics deviations
possible in Higgs data.

\end{abstract}

%\pacs{???}

\maketitle

%%%%%%%%%%%%%%%%%%%%%%%%%%%%%
%%%%%%%%%%%%%%%%%%%%%%%%%%%%%
\section{Introduction}
\label{sec:Introduction}

While the importance of the discovery of the Standard Model (SM)-like
Higgs boson by the CMS and ATLAS experiments~\cite{CMS:ril,
  Chatrchyan:2013mxa, CMS:bxa, ATLAS:2013oma, ATLAS:2013nma,
  ATLAS:2013wla} cannot be understated, many experimental tests remain
in order to determine whether the new particle is truly the Standard
Model Higgs boson.  Correspondingly, searches for non-Standard
Model-like properties of the new particle are highly motivated,
especially since a natural solution to the gauge hierarchy problem
invariably leads to deviations in the couplings of the Higgs boson to
SM states~\cite{Dawson:2013bba}.  So far, coupling extractions done by
both theorists~\cite{Azatov:2012bz, Espinosa:2012ir, Azatov:2012rd,
  Klute:2012pu, Carmi:2012zd, Blum:2012ii, Dobrescu:2012td,
  Carmi:2012in, Plehn:2012iz, LHCHiggsCrossSectionWorkingGroup:2012nn,
  Corbett:2012ja, Belanger:2013kya, Falkowski:2013dza,
  Giardino:2013bma, Djouadi:2013qya, Corbett:2013hia,
  Artoisenet:2013puc, Pomarol:2013zra, Boos:2013mqa, Stal:2013hwa} and
experimentalists~\cite{ATLAS:2013mma, ATLAS:2013sla, Aad:2013wqa,
  CMS:yva, ATLAS-CONF-2014-010} have concluded overall consistency
with a purely SM Higgs boson.  Further studies of the spin
properties~\cite{ATLAS:2013mla, Aad:2013xqa, ATLAS:2013xla,
  ATLAS:2013vla, Chatrchyan:2013iaa, CMS:2013wda} have corroborated
the SM-like nature of the new boson.  Nevertheless, these coupling
fits admittedly require model assumptions about how separate analyses
in different Higgs decay states are related.  Crucially, the SM
production mechanisms in these coupling fits are only deviated by
changing their respective signal strengths.  Our goal is to explore
exotic production of the SM-like Higgs boson, characterize how exotic
production modes can be differentiated from SM production modes, and
quantify the extent to which current Higgs analyses can be
contaminated by a new exotic production mode benchmark.

Our work strengthens the current and future of the LHC effort in
extracting as much information as possible from the Higgs signal,
including the Higgs discovery analyses from ATLAS~\cite{ATLAS:2013oma,
  ATLAS:2013nma, ATLAS:2013wla, ATLAS:tautau} and CMS~\cite{CMS:ril,
  Chatrchyan:2013mxa, CMS:bxa, Chatrchyan:2014nva}.  We provide an
initial foray into the direct study of exotic production using the
current Higgs analyses and are thus distinct from proposed
improvements of Higgs production tests~\cite{Li:2013ava,
  Rentala:2013uaa, Isidori:2013cga, Banerjee:2013apa, Gao:2013nga,
  Azatov:2013xha, Englert:2013vua, Maltoni:2013sma}.  It is
complementary to recent work on exotic decays of the Higgs
boson~\cite{Huang:2013ima, Curtin:2013fra}, which already have
experimental limits in a small number of channels, most notably the
possible invisible decay of the Higgs boson to dark matter
candidates~\cite{ATLAS:2013pma, CMS:1900fga, CMS:2013bfa,
  CMS:2013yda}.  In contrast, while probes of new exotic decays
typically require dedicated analyses, exotic production modes of the
Higgs are probed by kinematic distributions of the Higgs candidate and
the other particles produced in association with the Higgs.  These
studies can be done with current data and minimal modification to
current analyses and may reveal hints of a new exotic production mode
for the Higgs boson.  Such studies are vitally important in testing
whether Higgses at the LHC arise only from SM production modes.

In this letter, we will demonstrate the viability and importance of
models featuring exotic Higgs production.  We will first discuss the
general phenomenology of exotic production in~\secref{general}.  We
will then present benchmark models of chargino-neutralino production
in the minimal supersymmetric standard model (MSSM), and discuss its
phenomenology and the current experimental status
in~\secref{chi1chi2}.  In~\secref{effects}, we analyze how the new
exotic mode will affect the Higgs discovery analyses and discuss the
tests needed to disentangle SM and exotic production modes.  We
conclude in~\secref{conclusion}.  Auxilliary material derived from the
Higgs discovery analyses is in~\secref{expeffs}.

%%%%%%%%%%%%%%%%%%%%%%%%%%%%%
%%%%%%%%%%%%%%%%%%%%%%%%%%%%%
\section{Phenomenology of exotic production of the Higgs boson}
\label{sec:general}

The number of Higgs candidates in any given final state scales as
\begin{equation}
N_{\text{events}} = L \sigma \text{Br } \epsilon \sim \frac{g_p^2
  g_d^2}{\Gamma} \ ,
\label{eqn:rates}
\end{equation}
where $L$ is the integrated luminosity, $\sigma$ is the production
cross section, Br is the appropriate branching fraction, $\epsilon$ is
the signal efficiency, $g_p$ is the production coupling of the Higgs,
$g_d$ is the decay coupling of the Higgs, and $\Gamma$ is the total
Higgs width.  For the Higgs resonance, since the Higgs width is too
narrow to be identified directly\footnote{Indirect extraction of the
  Higgs width has been proposed, explored, and recently performed in
  Refs.~\cite{Caola:2013yja, Campbell:2013una, Campbell:2013wga,
    CMS:2014ala}.}, each final state only determines the combination
of $g_p^2 g_d^2 / \Gamma$.  Moreover, the production mechanism cannot
be tuned at the LHC, and so multiple signal production modes typically
contribute to a given final state analysis.  The contamination from
multiple production modes is well-known and can be mitigated by strict
cuts.  For example, the 2-jet category of the diphoton analysis is
most suited to vector boson fusion production but can have 20\% to
50\% contamination from gluon fusion production~\cite{ATLAS:2013oma,
  CMS:ril}.  Knowing the efficiencies of how different production
modes contribute to a given final state is critical in order to
appropriately combine separate rate measurements and extract
production and decay couplings.  Thus far, exotic production modes
have been neglected in these coupling scans, leaving open the
possibility that current coupling fits are overinterpreting the Higgs
dataset.  We will demonstrate that exotic production modes are viable
and characterize how their signatures differ from SM production modes.

Exotic production modes for the SM-like Higgs boson can arise from new
physics models with heavy particles whose decay products include the
126 GeV Higgs.  Familiar signatures of this type include new
resonances in two Higgs doublet models, where the heavy $CP$-even
Higgs scalar $H$ can decay to a pair of light $CP$-even Higgses, $H
\to hh$, or the $CP$-odd pseudoscalar $A$ can decay to the SM $Z$
boson and the light Higgs boson, $A \to Zh$.  These have been recently
searched for at CMS~\cite{CMS:2013eua}, excluding $\sim$few pb $\sigma
\times \text{Br}$ for resonances in the 260 to 360 GeV region.
% 8.5 to 5 pb for $H$ and 1.7 to 1.2 pb for $A$ on $\sigma \times $ Br.

Alternatively, new particles can produce cascade decays that could
include Higgs bosons.  Stops and gluinos in the MSSM, for example, can
cascade decay to heavy neutralinos, which can then decay to Higgs
bosons and lighter neutralinos.  Since these parent particles can be
strongly produced, the MSSM rate for Higgs production could have been
very large and exceed SM production rates.  Moreover, these cascade
decays of heavy superpartners have interesting phenomenological
signals~\cite{Datta:2003iz, Huitu:2008sa, Gori:2011hj, Stal:2011cz},
including boosted Higgs signatures for the $b \bar{b}$
channel~\cite{Butterworth:2008iy, Kribs:2009yh, Kribs:2010hp}.  The
key players for exotic Higgs production in the MSSM, though, are
typically the heavy neutralinos, and thus the strong production rates
studied above are absent if new colored states are too heavy to be
produced, as is currently favored by present data.  Nevertheless,
observable signals with weak scale cross sections are still
interesting.  Correspondingly, Drell-Yan chargino-neutralino
production has been extensively studied~\cite{Gunion:1987kg,
  Gunion:1987yh, Bartl:1988cn, Djouadi:2001fa, Diaz:2004qt} and
recently revisited~\cite{Baer:2012ts, Ghosh:2012mc, Arbey:2012fa,
  Baer:2013vpa, Baer:2013ava, Baer:2013faa, Han:2013mga, Baer:2013ssa,
  Berggren:2013vfa, Bharucha:2013epa, Gori:2013ala, Batell:2013bka,
  Han:2013kza, Berggren:2013bua, Buckley:2013kua,
  Papaefstathiou:2014oja}.

The cascade decay class of exotic production can be divided further
into symmetric and asymmetric subcategories.  Decay cascades giving
Higgses from symmetric production, such as stop--anti-stop production
from the MSSM, necessarily have a predicted correlation between
single- and double-Higgs production.  Namely,
\begin{eqnarray}
\sigma_{\text{h+X}} &=& \sigma \times 2 \mathcal{B} \left( 1 -
\mathcal{B} \right) \ , \\ 
\sigma_{\text{hh+X'}} &=& \sigma \times \mathcal{B}^2 \ ,
\end{eqnarray}
where $\sigma$ is the total cross section and $\mathcal{B}$ is the
overall decay chain branching fraction to a SM-like Higgs, assuming
the decay chain does not produce multiple Higgs particles.  This
correlation implies strong constraints on symmetric exotic production
models driven by current limits on double Higgs production.  The
current best searches for double Higgs production rely on the $b
\bar{b} \gamma \gamma$ final state~\cite{Chatrchyan:2013mya}, while
the highest significance Higgs decay modes are the $\gamma \gamma$,
$ZZ^* \to 4 \ell$, and $WW^* \to \ell \nu \ell \nu$ analyses.  The
correlated rates for an exotic symmetric production mode in the single
Higgs $\gamma \gamma$ final state with the double Higgs $b \bar{b} +
\gamma \gamma$ final state are then
\begin{eqnarray}
\sigma_{\text{h+X}} &=& \sigma \times 2 \mathcal{B} 
\left( 1 - \mathcal{B} \right) \times \text{Br}(h \to \gamma \gamma) \ , \\
\sigma_{\text{hh+X'}} &=& \sigma \times \mathcal{B}^2 \times
2 \text{Br}(h \to \gamma \gamma) \text{Br}(h \to b \bar{b})
\ ,
\end{eqnarray}
giving a single- to double-Higgs ratio $\frac{1- \mathcal{B}}{
  \mathcal{B} \times \text{ Br}(h \to b \bar{b})} \sim \frac{1 -
  \mathcal{B}}{0.6 \mathcal{B}}$.  For large values of $\mathcal{B}$,
the effect on double Higgs rates would be larger than the added single
Higgs rate, implying searches for double Higgs production are a better
probe of such scenarios.  Conversely, for small $\mathcal{B}$, the
large non-Higgs decay rate implies the main search modes for such
models would not be via Higgs final states.  Of course, double Higgs
analyses are insensitive to asymmetric cascade decay exotic production
modes, and so asymmetric parent particle production is a particularly
motivated class of exotic Higgs production.

Exotic production modes for the Higgs-like scalar exist in many beyond
the standard model theories.  Previous studies, however, largely
neglected the regions of parameter space where the exotic production
kinematics overlapped with SM production kinematics (an exception,
though, is~\cite{Howe:2012xe}).  As we enter the era of
precision studies of Higgs properties, it is thus timely and important
to understand these previously unexplored regions of parameter space.
Quantifying the extent of possible overlap between exotic and SM
production will be a main result of this work.

Higgs production modes are distinguished by two classes of
measurements: (1) kinematics of the Higgs candidate, and (2)
kinematics and multiplicites of the objects produced in association
with the Higgs.  This is readily justified because of the narrow Higgs
width and the scalar-like nature favored from spin studies, which
allow the Higgs decay state kinematics to be calculated in the Higgs
rest frame and then boosted according to the production mode
kinematics.  Inclusive Higgs rates, since they sum over contributions
of several different production modes with different efficiencies,
cannot disentangle the precise contribution of each contributing mode.
On the other hand, exclusive measurements exercise binning by
multiplicities of associated leptons and/or jets in the event to
isolate particular production modes.  Binning by jet and lepton
multiplicity is already in use by the experiments to attempt to
disentangle vector boson fusion (VBF), vector boson association (VBA)
and gluon fusion (ggF) production modes.  Yet, this is insufficient if
new physics (NP) modes are also present, since these rates can only
determine combinations of sums of production modes and their
respective efficiencies.  Different production modes are instead best
tested using differential distributions of the Higgs candidates and
the associated objects.  These distributions, such as Higgs $p_T$ and
$\eta$, associated jet and lepton $p_T$ and the event-level missing
transverse energy (MET), readily constrain the rates and shapes of all
production modes present in a given final state analysis.  Cross
correlations between orthogonal final states in related differential
distributions will then test whether the production profile of the
Higgs in one channel is consistent with other channels.

To this end, we motivate the simultaneous study of both rate
information and differential distributions in order to anatomize the
Higgs production modes.  We will focus now quantifying how measured
rates, which forms the basis for Higgs coupling fits, can be
contaminated by additional exotic production modes.

%%%%%%%%%%%%%%%%%%%%%%%%%%%%%
\subsection{Higgs rates and exotic production}
\label{subsec:rates}

In the Standard Model, the main production mode for the Higgs boson at
hadron colliders is gluon fusion, which dominates over the remaining
vector boson fusion, $W^\pm h$, $Zh$, and $t\bar{t}h$ production
modes.  In this vein, additional subleading SM production modes, such
as $th$ or triple boson production including a Higgs, are neglected
and, especially if they are enhanced by new physics, can be considered
as exotic production modes.  Moreover, while direct searches for new
physics at the LHC have excluded some large regions of NP parameter
space, only a limited number of direct searches consider final states
that include Higgs bosons.  As the Higgs boson may be the leading
connection between SM and new physics sectors, it is worthwhile to
consider that new physics first shows up in Higgs physics instead of
the traditional dedicated NP analyses.  Clearly, NP scenarios can thus
be constrained from studies of Higgs rates: however, as mentioned
before, such rates only constrain linear sums of various production
modes and are thus not truly model-independent.  Moreover, it is
possible that SM production rates are reduced via NP
effects~\cite{Kumar:2012ww}, but exotic production is simultaneously
present to make up any noticeable difference.

We can subdivide NP exotic production rates into three categories: (A)
$\sigma_{NP} \gtrsim \sigma_{SM}$, (B) $\sigma_{NP} \sim \sigma_{SM}$,
and (C) $\sigma_{NP} \lesssim \sigma_{SM}$, where $\sigma_{SM}$ refers
to the various calculated SM Higgs production cross sections.  The
high resolution $\gamma \gamma$ and $ZZ^* \to 4 \ell$ channels from
ATLAS~\cite{ATLAS:2013oma, ATLAS:2013nma, ATLAS:2013mma,
  ATLAS:2013sla, Aad:2013wqa} and CMS~\cite{CMS:ril,
  Chatrchyan:2013mxa, CMS:yva} give combined Higgs mass measurements
of 125.5 $\pm$ 0.2 (stat.)$^{+0.5}_{-0.6}$ (syst.) GeV [ATLAS] and
125.7 $\pm$ 0.3 (stat.) $\pm$ 0.3 (syst.) GeV [CMS].  Approximating
the Higgs mass as 126 GeV, we list the dominant SM production modes
and their percentage uncertainties in~\tableref{prod_xsecs}, taken
from Refs.~\cite{Dittmaier:2011ti, Dittmaier:2012vm,
  Heinemeyer:2013tqa}.

\begin{table}[tbh]
\renewcommand{\arraystretch}{1.2}
\begin{tabular}{|c|c|c|c|}
\hline
Mode & Cross section (pb) & QCD scale (\%) & PDF+$\alpha_s$ (\%) \\
\hline
ggF & 18.97 & (+7.2, -7.8) & (+7.5, -6.9) \\
VBF & 1.568 & (+0.3, -0.1) & (+2.6, -2.8) \\
$W^\pm h$ & 0.6860 & (+1.0, -1.0) & (+2.3, -2.3) \\
$Zh$ & 0.4050 & (+3.2, -3.2) & (+2.5, -2.5) \\
$t\bar{t} h$ & 0.1262 & (+3.8, -9.3) & (+8.1, -8.1) \\
\hline
\end{tabular}
\caption{From Ref.~\cite{Heinemeyer:2013tqa}, SM Higgs production
  cross sections and theoretical uncertainties for the main production
  modes for 126 GeV Higgs at 8 TeV LHC.}
\label{table:prod_xsecs}
\end{table}

Since current Higgs rates are roughly consistent with SM expectations,
we find only category (B) is worthwhile for detailed study at present.
Category (A) is viable if the NP signal efficiency is tiny, which
requires very non-standard kinematics (such as displaced vertices
giving Higgs decays~\cite{Jaiswal:2013xra}), which would be better
addressed by dedicated searches.  Category (C) is also presently
unmotivated, since the current experimental and theoretical
uncertainties on Higgs rates are too large to possibly constrain
additional exotic production modes that are subleading compared to SM
rates.  For category (B), knowing the signal efficiencies for various
SM and NP production modes is key to understanding how sensitive rate
analyses will be in disentangling NP from SM production, which is the
focus of~\secref{effects}.

We could include another category of NP Higgs exotica, namely exotic
decays of the Higgs boson (see, e.g. Refs.~\cite{Huang:2013ima,
  Curtin:2013fra}).  We view the possibilities of exotic production
modes and exotic decays of the Higgs, however, as complementary probes
for understanding the connections between the Higgs and new physics.
More practically, the phenomenology of exotic production and exotic
decays require different tools.  Probing exotic production requires a
well-defined Higgs candidate and then analyzing the global event
properties to test how it was produced.  Exotic decays of the SM-like
Higgs, however, require dedicated searches for new final states, where
the exotic Higgs signal candidates must be isolated from background.
In this work, we will only focus on exotic production, though
combining exotic production and exotic decay signatures would be
interesting for future study.  Thus, we will assume the SM Higgs
partial decay widths are not significantly modified by NP and that no
new decay modes for the Higgs are introduced.

%%%%%%%%%%%%%%%%%%%%%%%%%%%%%
%%%%%%%%%%%%%%%%%%%%%%%%%%%%%
\subsection{Current sensitivity to exotic production}
\label{sec:sensitivity}

The experiments have done several searches for Higgs signatures,
aiming to probe various final states and, to a separate degree,
isolating particular production channels.  We first discuss the
overall status of Higgs measurements and then isolate the most
relevant Higgs analyses that are sensitive to exotic production modes.
We outline the searches below, critically evaluating each search and
its sensitivity to exotic production signals.

The extent to which a new exotic production mode is categorized in the
current experimental searches is best understood by differentiating
exclusive and inclusive Higgs production searches.  More inclusive
Higgs searches, such as the diphoton, fully leptonic $ZZ^*$ channel,
fully leptonic $WW^*$ channel, and $\tau \tau$ analyses from
ATLAS~\cite{ATLAS:2013oma, ATLAS:2013nma, ATLAS:2013wla, ATLAS:tautau}
and CMS~\cite{CMS:ril, Chatrchyan:2013mxa, CMS:bxa,
  Chatrchyan:2014nva} use cuts on jets, leptons, and missing energy in
order to partition the contributions of separate SM production modes.
Other searches singly target vector boson fusion
production~\cite{CMS:2013yea, CMS:2013jda}, vector boson associated
production~\cite{CMS:ckv, CMS:zwa, Chatrchyan:2013zna, CMS:2013xda,
  TheATLAScollaboration:2013hia, TheATLAScollaboration:2013lia}, or $t
\bar{t} h$ production~\cite{CMS:2013sea, CMS:2013fda, CMS:2013tfa,
  TheATLAScollaboration:2013mia}.  While each result standalone can
readily be interpreted as an observed experimental rate, the
combination of separate searches relies crucially on assumptions about
the underlying production modes and is hence altered when a new exotic
mode is considered.  As mentioned before, the combinations from
ATLAS~\cite{ATLAS:2013mma, ATLAS:2013sla, Aad:2013wqa} and
CMS~\cite{CMS:yva} do not consider possible contamination of the Higgs
dataset from exotic production modes.  In order to determine the
extent and constraints on new exotic production modes, the rates,
kinematics and multiplicities of associated objects, and the
kinematics of the Higgs candidate must all be measured and given.  In
addition, the efficiencies for the exotic production mode must be
simulated in order to know how the new mode is tagged and possibly
disentangled from the SM modes.

In principle, any new putative exotic Higgs production mode must be
considered in all possible Higgs decay final states.  The most
important analyses for our purposes, however, are the $\gamma \gamma$,
$ZZ^* \to 4 \ell$, and $WW^* \to \ell \nu \ell \nu$ measurements,
since these analyses make the most use of event categorization based
on objects produced in association with the Higgs.  Other final
states, such as $\tau \tau$ and $b \bar{b}$, have less sensitivity to
the SM Higgs production rates with current data, and are expected to
be less powerful in disentangling new production modes unless the
exotic mode itself is rich in $\tau$ or $b$ production.  Singly
targeted analyses for SM VBF, VBA, or $t \bar{t} h$ production, if
they do not categorize according to expected SM production
contributions, each measure only a single rate.  Hence, their impact
beyond the discovery modes is minimal, since the expected cross
correlations from a new exotic production mode in separate categories
should already be present in the discovery modes.  Thus, we focus our
attention on the Higgs discovery modes of $\gamma \gamma$, $ZZ^* \to 4
\ell$, and $WW^* \to \ell^+ \nu \ell^- \nu$ final states.  These have
the additional advantage of being very cleanly measured final states,
and so further analyses of differential distributions are highly
promising and well-motivated\footnote{Also advocated in
  Ref.~\cite{Boudjema:2013qla}.}.  While this has been done by ATLAS
in the $\gamma \gamma$ channel~\cite{TheATLAScollaboration:2013eia},
we look forward to future presentations of differential distributions
by ATLAS and CMS, possibly including the $4 \ell$ final state.

We can see that a well-motivated and timely scenario for further study
is characterized by cross sections similar to SM production rates and
regions of parameter space where the exotically produced Higgs has
kinematics similar to SM kinematics.  Moreover, to readily avoid
constraints on double Higgs production, it is simplest to consider
asymmetric production of parent particles that cascade decay to a
Higgs in the final state.  Surveying possible exotic production modes,
we now consider chargino-neutralino production in the MSSM as a
well-motivated benchmark for studying exotic production.

%%%%%%%%%%%%%%%%%%%%%%%%%%%%%
%%%%%%%%%%%%%%%%%%%%%%%%%%%%%
\section{Chargino-neutralino production in the MSSM}
\label{sec:chi1chi2}

In the minimal supersymmetric standard model, there are many possible
cascade decay chains that produce Higgs bosons.  Squarks, for example,
can decay to Higgs bosons via flavor-conserving heavy to light flavor
decays.  Neutralinos and charginos, via mixing, can decay to Higgs
bosons in transitions from heavy to light mass eigenstates.  We will
focus on the cascade decay of a heavy neutralino into a light
neutralino and the SM-like Higgs boson.

\begin{figure}[htb]
\begin{center}
\includegraphics[scale=0.7, angle=0]{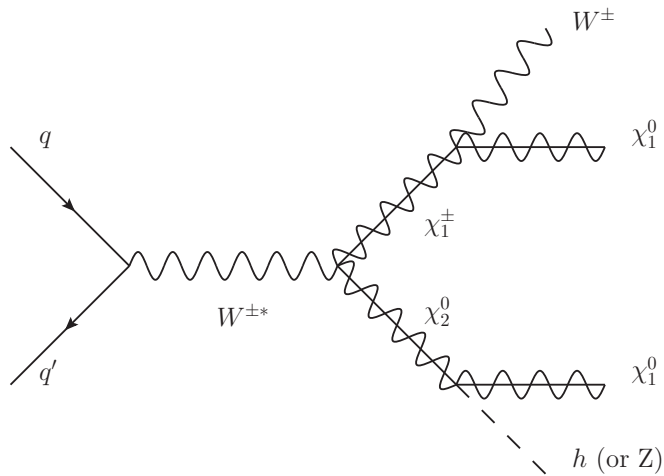}
\caption{\label{fig:feynman} Feynman diagram for $\chi_1^{\pm} \chi_2^0$
    production.}
\end{center}
\end{figure}

Chargino-neutralino production at the LHC dominantly proceeds via an
$s$-channel $W^\pm$ boson, and assuming simple two-body decays of the
chargino and neutralino, leads to the final state of $W^\pm h \chi_1^0
\chi_1^0$ or $W^\pm Z \chi_1^0 \chi_1^0$, as shown
in~\figref{feynman}.  If the $W^\pm h$ cascade decay is favored over
the $W^\pm Z$ cascade decay, then the $\chi_1^\pm \chi_2^0$ production
will not be seen in traditional multilepton searches and will instead
be probed by Higgs exotic production searches.  The important
parameters in determining these relative rates are the electroweak
gaugino mass terms, $M_1$, $M_2$, $\mu$, and $\tan \beta$.  The first
three, apart from potentially significant mixing effects, determine
the bino, wino, and higgsino masses.  After $SU(2)$ breaking, the
bino, neutral wino, and two neutral higgsino states mix, and the
charged wino and charged higgsino mix.

%%%%%%%%%%%%%%%%%%%%%%%%%%%%%
%%%%%%%%%%%%%%%%%%%%%%%%%%%%%

The tree level interactions between neutral gauginos in the gauge
basis are easy to identify.  There are diagonal $Z$ mediated
interactions for $\tilde{H}_d^0$ and also for $\tilde{H}_u^0$.  After
mixing, the particular combination of gauge eigenstates given by
$\frac{1}{\sqrt{2}} \left( \tilde{H}_d^0 \pm \tilde{H}_u^0 \right)$
can exchange $Z$ bosons between each other but do not have diagonal
$Z$ couplings.  Separately, Higgs exchange only exists between
Higgsino eigenstates and gaugino eigenstates.  So, while pure neutral
winos and pure binos have no tree-level vertices amongst themselves,
wino-like neutralinos can decay to Higgses and bino-like neutralinos
via a Higgsino mixing angle insertion, while a $Z$-mediated decay
requires two Higgsino mixing angles.  Because of possible intrinsic
cancellations among up and down Higgsino mixing angles, however, the
$Z$-mediated decay may still dominate.  A more extensive discussion of
the relative branching fractions of next-to-lightest supersymmetric
particles (NLSPs) is available in Ref.~\cite{Han:2013kza}, but we can
simplify the parameter space to variations in $\mu$ and $M_2$, holding
$M_1$ and $\tan \beta$ fixed.

\begin{figure}[htb]
\begin{center}
\psfrag{xaxislabel1234567890}{$M_2$ (GeV)}
\psfrag{yaxislabel1234567890}{Br$(\chi_2^0 \to X \chi_1^0)$}
\psfrag{topaxislabel123456789012345678901234567890}{Wino-like NLSP, $\mu = 800$ GeV}
\includegraphics[scale=0.75, angle=0]{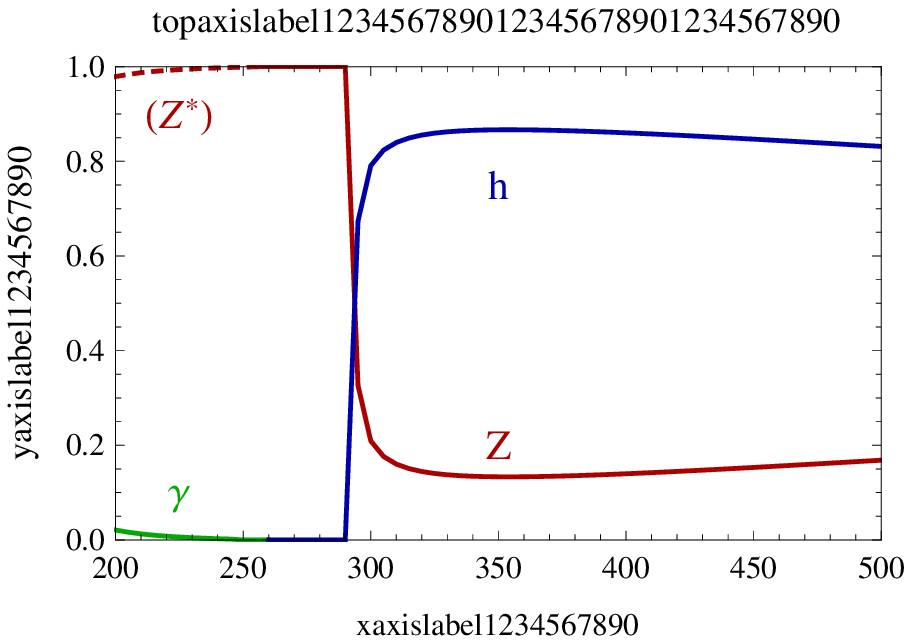}
\quad
\psfrag{xaxislabel1234567890}{$M_2$ (GeV)}
\psfrag{yaxislabel1234567890}{Br$(\chi_2^0 \to X \chi_1^0)$}
\psfrag{topaxislabel123456789012345678901234567890}{Wino-like NLSP, $\mu = 2000$ GeV}
\includegraphics[scale=0.75, angle=0]{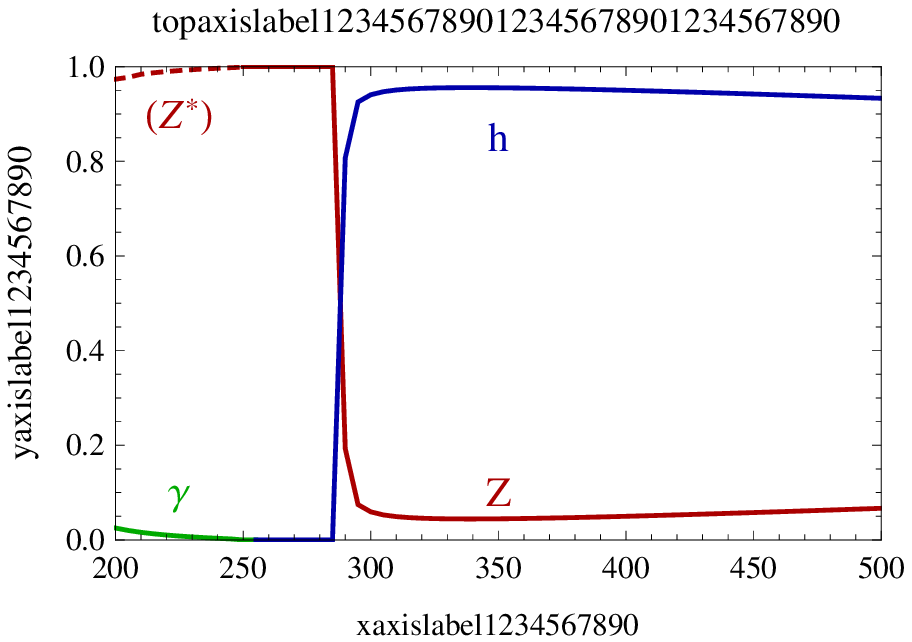}

\vspace{0.2in}
\psfrag{xaxislabel1234567890}{$\mu$ (GeV)}
\psfrag{yaxislabel1234567890}{Br$(\chi_2^0 \to X \chi_1^0)$}
\psfrag{topaxislabel123456789012345678901234567890}{Higgsino-like NLSP, $M_2 = 500$ GeV}
\includegraphics[scale=0.75, angle=0]{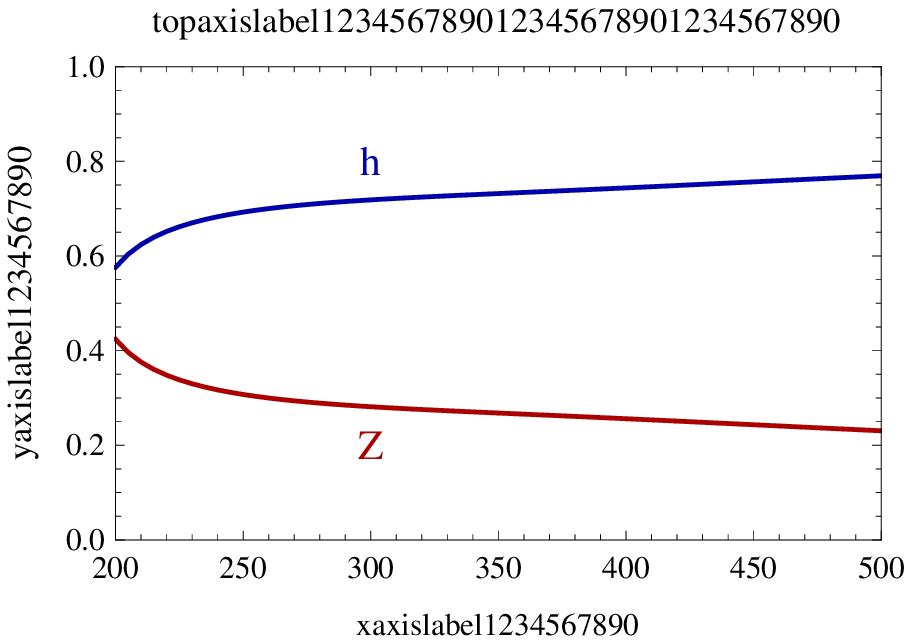}
\quad
\psfrag{xaxislabel1234567890}{$\mu$ (GeV)}
\psfrag{yaxislabel1234567890}{Br$(\chi_2^0 \to X \chi_1^0)$}
\psfrag{topaxislabel123456789012345678901234567890}{Higgsino-like NLSP, $M_2 = 2000$ GeV}
\includegraphics[scale=0.75, angle=0]{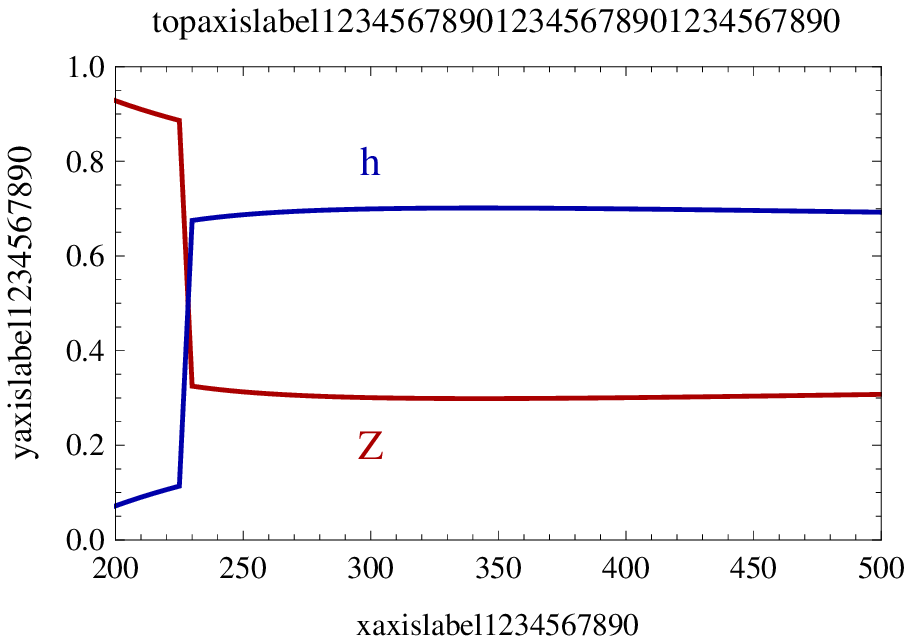}
\caption{\label{fig:chi20BRs} Relevant branching fractions of
  $\chi_2^0 \to h \chi_1^0$ (blue), $Z \chi_1^0$ (red), $\gamma
  \chi_1^0$ (green).  Each panel uses SUSY soft mass parameters $M_0$
  = 2 TeV, $\tan \beta = 10$, $m_A = 2$ TeV, $A_0 = 2.5$ TeV, $M_1 =
  200$ GeV, $M_3 = 2$ TeV at the mSUGRA scale.  }
\end{center}
\end{figure}

In~\figref{chi20BRs}, we show that Br$(\chi_2^0 \to h \chi_1^0)$ can
readily dominate over Br$(\chi_2^0 \to Z \chi_1^0)$.  Spectra are
calculated with SUSY soft mass parameters $M_0 = 2$ TeV, $\tan \beta =
10$, $m_A = 2$ TeV, $A_0 = 2.5$ TeV, $M_1 = 200$ GeV, $M_3 = 2$ TeV
using~\textsc{SOFTSUSY v3.3.8}~\cite{Allanach:2001kg}, and decay
tables are calculated with \textsc{SUSYHIT v1.3}~\cite{Djouadi:2006bz,
  Djouadi:2002ze, Djouadi:1997yw, Muhlleitner:2003vg}.  The upper
panels have wino-like NLSPs, with $M_2 < \mu$.  Because of mixing with
the bino, the mass splitting between the NLSP and LSP is not large for
small $M_2$, but once the mass splitting exceeds the Higgs threshold,
the branching fraction for $\chi_2^0 \to h \chi_1^0$ easily dominates
over $Z \chi_1^0$.  Similarly, the bottom panels have Higgsino-like
NLSPs, with $\mu \lesssim M_2$ in the lower left panel and $\mu < M_2$
in the lower right panel.  Since the Higgs decay channel is
kinematically open for this entire parameter range, we see from the
lower left panel that it is favored compared to the $Z$ channel.  The
lower right panel has the same behavior except for low $\mu$, where a
level-crossing occurs and the different sign in the $\chi_2^0$ gauge
eigenstate composition leads a dominant $Z$ decay to the LSP.  We can
see the complementary nature of exotic production probes and
multilepton searches in this scenario, whose sensitivities are driven
by Br($\chi_2^0 \to h \chi_1^0$) and Br($\chi_2^0 \to Z \chi_1^0$).
For all of these points, the associated chargino $\chi_1^+$ is nearly
mass-degenerate with $\chi_2^0$ and decays to $W^+ \chi_1^0$
(sometimes via an off-shell $W^*$) 100\% of the time.

%%%%%%%%%%%%%%%%%%%%%%%%%%%%%
%%%%%%%%%%%%%%%%%%%%%%%%%%%%%

\begin{figure}[htb]
\begin{center}
\psfrag{x1234567890123456789012345678901234567890}{$\mu$ (red, blue) or $M_2$ (green, purple) (GeV)}
\psfrag{y123456789012345678901234567890}{NLO $\sigma(pp \to \chi_1^\pm \chi_2^0)$ (pb) for 8 TeV LHC}
\psfrag{A}{{\bf Model A}}
\psfrag{B}{{\bf Model B}}
\psfrag{C}{$\blacktriangle$}
\psfrag{D}{$\blacksquare$}
\includegraphics[scale=1.2, angle=0]{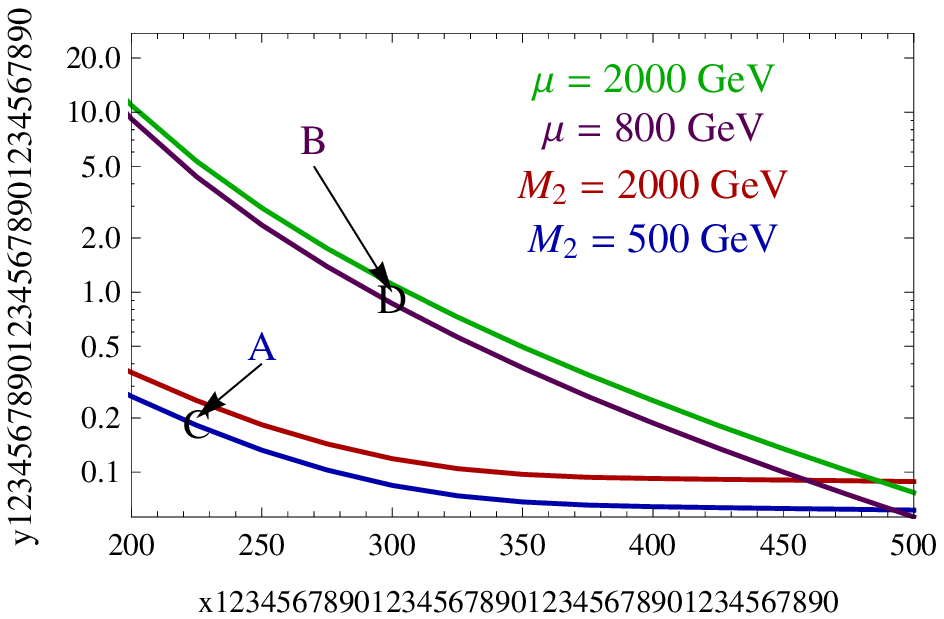}
\caption{\label{fig:xsecs} Cross sections for $pp \to \chi_1^\pm
  \chi_2^0$ (pb) at 8 TeV LHC as labeled.  Note the horizontal axis is
  either $\mu$ or $M_2$ according to the relevant curve.}
\end{center}
\end{figure}

In addition, the production cross sections for $pp \to \chi_1^\pm
\chi_2^0$ can be large, which we show in~\figref{xsecs} as a function
of the same mass parameters as before: either $M_2$ ($\mu$ fixed to
800 GeV or 2 TeV) or $\mu$ ($M_2$ fixed to 500 or 2 TeV).  These cross
sections are calculated using \textsc{Prospino
  v.2.1}~\cite{Beenakker:1999xh}.  Recall that $Wh$ production in the
SM is 0.6860 pb (for $m_h = 126$ GeV), so the contamination fraction
for $Wh + $ MET relative to $Wh$ in the SM can be order 1 or larger.

%%%%%%%%%%%%%%%%%%%%%%%%%%%%%
%%%%%%%%%%%%%%%%%%%%%%%%%%%%%

%%%%%%%%%%%%%%%%%%%%%%%%%%%%%
\subsection{Benchmark models for exotic production}
\label{subsec:benchmarks}

We will study higgsino-like NLSPs and wino-like NLSPs, keeping in mind
from~\figref{chi20BRs} that multilepton searches give indirect
constraints on these exotic Higgs production models.  We choose
benchmark points as shown in~\tableref{benchmarks}.  The mass spectra
are calculated using~\textsc{SOFTSUSY v3.3.8}~\cite{Allanach:2001kg},
and branching fractions are calculated with \textsc{SUSYHIT
  v1.3}~\cite{Djouadi:2006bz, Djouadi:2002ze, Djouadi:1997yw,
  Muhlleitner:2003vg}.  We use \textsc{Prospino
  v.2.1}~\cite{Beenakker:1999xh} for the next-to-leading order (NLO)
cross sections at 8 TeV LHC, and relic abundance is calculated using
\textsc{micrOMEGAs v.3.6.7}~\cite{Belanger:2013oya}.  We now detail
the model phenomenology and current constraints from direct searches,
multilepton studies, and dedicated $Wh + $ MET searches.

\begin{table}[tbh]
\renewcommand{\arraystretch}{1.2}
\begin{tabular}{|c|c|c|}
\hline
{\bf Parameter} & {\bf Model A} & {\bf Model B} \\
\hline
$M_0$ & 2000 & 2000 \\
$\tan \beta$ & 10 & 10 \\
$m_A$ & 2000 & 2000 \\
trilinear $A_0$ & 2500 & 2500 \\
$M_1$ & 200 & 200 \\
$M_2$ & 500 & 300 \\
$M_3$ & 2000 & 2000 \\
$\mu$ & 225 & 800 \\
\hline
$\chi_1^\pm$ mass & 213 GeV & 191 GeV \\
$\chi_2^0$ mass & 215 GeV & 191 GeV \\
$\chi_1^0$ mass & 57.8 GeV & 61.5 GeV \\
\hline
BR($\chi_2^0 \to h \chi_1^0$) & 66.2\% & 79.1\% \\
BR($\chi_2^0 \to Z \chi_1^0$) & 33.8\% & 20.9\% \\
BR($\chi_2^\pm \to W^\pm \chi_1^0$) & 100\% & 100\% \\
\hline
$\sigma(\chi_1^+ \chi_2^0)$ & 0.126 & 0.622 \\
$\sigma(\chi_1^- \chi_2^0)$ & 0.058 & 0.295 \\
\hline
$\Omega h^2$ & 0.0211 & 0.117 \\
$\sigma_{SI, p}$ & $7.265 \times 10^{-10}$ & $2.193 \times 10^{-11}$ \\
$\sigma_{SD, p}$ & $5.858 \times 10^{-5}$ & $3.341 \times 10^{-7}$ \\
$\sigma_{SI, n}$ & $7.442 \times 10^{-10}$ & $2.251 \times 10^{-11}$ \\
$\sigma_{SD, n}$ & $4.488 \times 10^{-5}$ & $2.624 \times 10^{-7}$ \\
\hline
Br($h \to \chi_1^0 \chi_1^0$) & 0.035 & $3.60 \times 10^{-5}$ \\
\hline
\end{tabular}
\caption{Model parameters for benchmark models A and B.  Masses are in
  GeV, cross sections are in pb.}
\label{table:benchmarks}
\end{table}

First, the model benchmarks have heavy scalars and a gluino in the TeV
to multi-TeV range, which satisfy the limits on squarks and gluino
production in jets + MET final states.  Moreover, the radiative
corrections induced by the multi-TeV stops and gluino drive the
lightest SM-like Higgs mass to $125.5$ GeV within theory uncertainty.
Separately, the relic abundance of the bino-like LSP satisfies
constraints on $\Omega h^2 = 0.1198 \pm 0.0026$~\cite{Beringer:1900zz}
and the most recent direct detection constraints on dark
matter-nucleon cross sections~\cite{Akerib:2013tjd}, and so the LSP is
a suitable dark matter candidate.  Furthermore, the invisible decay of
$h \to \chi_1^0 \chi_1^0$ is suppressed below current constraints.
The remaining flavor constraints, direct searches for the heavy Higgs
sector and sleptons are also easily satisfied by the mass decoupling
of the spectra.  These two models, although fine-tuned, exemplify the
simplied model~\cite{Alves:2011wf} attitude.  Here, the LHC accessible
electroweak gaugino states and their interesting phenomenology can be
isolated and discussed separately from the model constraints on the
full theory, as demonstrated by explicit construction.  As a
corollary, this example of exotic production of the Higgs introduces
observable deviations in Higgs physics while staying within the MSSM
decoupling limit, which only constrains SM Higgs production and decay
rates!

We can see that Model A has a dominantly Higgsino-like
chargino-neutralino pair at about 215 GeV (in this model, $\chi_3^0$
has a mass of 239 GeV), while Model B has a dominantly Wino-like
chargino-neutralino pair at 191 GeV.  As previewed
in~\figsref{chi20BRs}{xsecs}, both scenarios have a suppressed $W Z +
$ MET cascade decay rate, which is probed in multilepton analyses
aimed at the leptonic decays of the gauge bosons.  Comparing to
multilepton event searches~\cite{Aaltonen:2013vca, Aad:2012xsa,
  Aad:2014nua, TheATLAScollaboration:2013hha, CMS:2013dea} and
rescaling the limits by the appropriate $\chi_2^0 \to Z \chi_1^0$
branching fraction, we see both benchmarks satisfy the current
multilepton constraints.  Moreover, direct chargino pair production
searches, interpreted without intermediate sleptons in the cascade
decay, are also satisfied, since the current limits are not yet
sensitive to these cross
sections~\cite{TheATLAScollaboration:2013hha}.

Separately, the $Wh + $ MET, $h \to bb$ or same-sign dilepton +
$jj(j)$ + MET searches~\cite{TheATLAScollaboration:2013zia,
  CMS:2013afa} are insensitive to our model benchmarks, since both
models lie relatively close to the Higgs mass difference line in the
$\chi_1^\pm/\chi_2^0$ mass vs. $\chi_1^0$ mass plane.  In these
searches, the expected SM $Wh$ contribution is cut from the analyses
via a hard transverse mass cut, rendering them insensitive to the
Higgs mass difference line where the cascade decay of the heavy
neutralino produces a very weakly boosted Higgs boson.  Hence, we can
additionally motivate our analyses of the Higgs discovery modes as
probing this complementary part of parameter space where the previous
$Wh + $ MET analyses do not have sensitivity.  The two benchmarks are
thus illustrative in understanding the effect of changing the
$\chi_2^0 - \chi_1^0$ mass difference in order to interpolate between
the above searches, which are sensitive to large mass splittings and
moderately boosted Higgses, versus the SM Higgs analyses, which are
probing the non-boosted regime of exotic production.

%%%%%%%%%%%%%%%%%%%%%%%%%%%%%
%%%%%%%%%%%%%%%%%%%%%%%%%%%%%
\section{Effects on current Higgs analyses from exotic production}
\label{sec:effects}

Having established our benchmark models, we proceed with quantifying
how they can be probed in the Higgs discovery analyses.  We perform
Monte Carlo simulations implementing the cut-and-count analyses of
ATLAS~\cite{ATLAS:2013oma, ATLAS:2013nma, ATLAS:2013wla} and
CMS~\cite{CMS:ril, Chatrchyan:2013mxa, CMS:bxa}.  This will determine
signal efficiencies for our benchmark exotic modes of the SM-like
Higgs.  We will then discuss the calculated efficiencies, make
comparisons to extrapolated SM signal efficiencies, and the necessary
probes to distinguish exotic and SM production modes.  We neglect the
effect of these exotic production benchmarks on the data-driven
background control regions and instead only focus on signal regions.

Event samples are generated using~\textsc{MadGraph5
  v1.5.7}~\cite{Alwall:2011uj} with the CTEQ6L1
PDFs~\cite{Pumplin:2002vw}.  Each event is passed through
\textsc{Pythia v6.4.20}~\cite{Sjostrand:2006za} for showering and
hadronization.  No detector simulation is used, but we do not expect
it to significantly change the results, because the final states are
high resolution channels (mainly photons and leptons) and the main
cuts of the analysis that define the signal region are driven by these
well-measured objects.

We use the cut and count analyses for the main Higgs discovery modes:
$h \to \gamma \gamma$, $h \to Z Z^* \to 4 \ell$, and $h \to W W^* \to
\ell \nu \ell \nu$, with $\ell = e$, $\mu$.  In
both~\tablesref{cutflowATLAS}{cutflowCMS}, the signal categorization
efficiencies are given relative to the Higgs rate to the specified
final state, except for the $WW^*$ efficiency, which uses the Higgs
rate $h \to W W^* \to (\ell + \tau) \nu (\ell + \tau) \nu$, to account
for the subleading contribution from leptonic decays of $\tau$s.
Numbers of expected events in the $\gamma \gamma$, for example, are
then given by $\sigma (pp \to \chi_1^\pm \chi_2^0) \times
\text{Br}(\chi_1^+ \to W^+ \chi_1^0) \times \text{Br}(\chi_2^0 \to h
\chi_1^0) \times \text{Br}_{SM}(h \to \gamma \gamma) \times
\epsilon_{\gamma \gamma}$, where $\epsilon_{\gamma \gamma}$ is the
appropriate category from the table.  By design, $\epsilon$ is the
same as defined in~\eqnref{rates}, whereby all Higgs branching
fractions are accounted for separately.  Note that since our
production mode has an extra $W$ boson, we generate two orthogonal
datasets for the $WW^*$ analysis: (1) $W \to $ anything, $h \to \ell
\nu \ell \nu$, and (2) $W \to \ell \nu$, $h \to \ell \nu j j$ with
$\ell = e$, $\mu$, or $\tau$, since each could contribution to the
$\ell \nu \ell \nu$ final state.  We combine the resulting separate
signal efficiencies, appropriately normalized to the $h \to (\ell +
\tau) \nu (\ell + \tau) \nu$ rate.  The ATLAS $\gamma \gamma$
high-mass two-jet category combines the original orthogonal ``loose''
and ``tight'' categorizations.  We only implement the gluon fusion
targeted $0-$jet and $1-$jet search from the original CMS $WW$
analysis, since the $2-$jet analysis has limited statistical
significance.

\begin{table}[tbh]
\renewcommand{\arraystretch}{1.2}
\begin{tabular}{|c|c|c|c|}
\hline
Analysis & Category                  & {\bf Model A} & {\bf Model B} \\
\hline
& Lepton               & 6.3\%  & 6.6\% \\
& MET significance     & 28.2\% & 22.7\% \\
$\gamma \gamma$ & Low-mass two-jet     & 1.4\%  & 1.9\% \\
& High-mass two-jet    & 0.2\%  & 0.2\% \\
& Untagged             & 9.1\% & 14.0\% \\
\hline
& ggF-like             & 21.5\% & 21.4\% \\
$ZZ^*$ & VBF-like             & 0.2\%  & 0.2\%  \\
& VH-like              & 7.1\%  & 7.1\% \\
\hline
& $N_{\text{jet}} = 0$    & 1.6\% & 1.7\% \\
$WW^*$ & $N_{\text{jet}} = 1$    & 3.4\% & 3.1\% \\
& $N_{\text{jet}} \geq 2$ & $<$0.1\% & $<$0.1\% \\
\hline
\end{tabular}
\caption{Signal categorization efficiencies for ATLAS $\gamma \gamma$,
  $ZZ^* \to 4 \ell$, and $WW^* \to \ell \nu \ell \nu$ analyses.  The
  $WW^*$ rows sum the two orthogonal signal generation samples of $Wh
  \chi_1^0 \chi_1^0$, $h \to \ell \nu \ell \nu$ ($\ell = e, \mu,
  \tau$), $W \to $ anything, and $Wh \chi_1^0 \chi_1^0$, $W \to \ell
  \nu$, $h \to \ell \nu j j$, as described in the main text.}
\label{table:cutflowATLAS}
\end{table}

While most of the analyses have included selection efficiency tables
for identification and reconstruction of signal objects, the ATLAS
$ZZ^*$ analysis only provides aggregate signal efficiencies.  These
are organized according to lepton final state and have expected signal
reconstruction efficiencies for a 125 GeV Higgs produced in the SM as
39\% for $4 \mu$, 26\% for $2e 2\mu$ and $2\mu 2e$, and 19\% for $4e$.
We incorporate these efficiencies into our analysis by simulating
gluon fusion production with a $h + 0/1 j$ matched sample and
calculating the appropriate lepton category-specific reweighting
factors.  We apply these reweighting factors to account for
identification and reconstruction efficiencies in our mock up of the
ATLAS $ZZ^*$ analysis, assuming these additional efficiencies only
depend on lepton category and have no dependence on kinematics.  As a
cross check of this ad hoc procedure, we see our exotic NP
efficiencies do mimic the extrapolated ATLAS SM $Wh + Zh$ efficiencies
shown~\tableref{ATLASZZ} of~\secref{expeffs}.  Namely, we see that the
ggF-like, VBF-like, and VH-like categorization of $Wh + Zh$ production
are 22.4\%, 0.3\%, and 4.8\%, respectively, which roughly follow our
calculated NP efficiencies.

\begin{table}[tbh]
\renewcommand{\arraystretch}{1.2}
\begin{tabular}{|c|c|c|c|}
\hline
Analysis & Category              & {\bf Model A} & {\bf Model B} \\
\hline
& Muon             & 5.2\% & 5.1\% \\
& Electron         & 5.1\% & 5.1\% \\
$\gamma \gamma$ & Dijet tight      & 0.1\% & 0.1\% \\
& Dijet loose      & 0.3\% & 0.3\% \\
& $E_T$ miss       & 26.7\% & 16.8\% \\
& Untagged         & 20.2\% & 32.5\% \\
\hline
$ZZ^*$ & Category 1, $N_{\text{jet}} \leq 1$ & 22.1\% & 22.9\% \\
& Category 2, $N_{\text{jet}} \geq 2$ & 11.6\% & 10.8\%  \\
\hline
$WW^*$ & $0-$jet             & 0.3\% & 0.4\% \\
& $1-$jet             & 1.0\% & 1.2\% \\
\hline
\end{tabular}
\caption{Signal categorization efficiencies for CMS $\gamma \gamma$,
  $ZZ^* \to 4 \ell$, and $WW^* \to \ell \nu \ell \nu$ analyses.
  $WW^*$ signal efficiencies are calculated the same
  as~\tableref{cutflowATLAS}.}
\label{table:cutflowCMS}
\end{table}

We can see that our two benchmarks have very similar signal
categorization efficiencies from the cut and count Higgs analyses,
with the notable exception of the MET categorization in the diphoton
analyses.  As expected, the larger mass splitting between $\chi_2^0$
and $\chi_1^0$ in Model A leads to a larger MET signal efficiency than
that for Model B.  With even heavier parent particles, the MET
efficiency will continue to grow and the Higgs would become more
boosted, giving us a smooth transition from these SM-focused analyses
and the dedicated boosted Higgs analyses.  If we kept the mass
splitting between the parent particles and the LSP constant and simply
raised all masses together, however, we expect these efficiency tables
to be largely unchanged.  These orthogonal directions in this
simplified model space can be explored simultaneously with a robust
two-dimensional efficiency matrix, which we reserve for future work.

We remark that besides the low (targeting hadronic $W$ and $Z$
candidates) and high (targeting VBF forward jets) dijet cuts and the
MET categorization, the remaining signal categories are simply
multiplicity bins, and hence do not probe the different kinematics of
the Higgs production mechanisms.  On the other hand, differential
distributions, especially in these well-measured, high-resolution
leptonic and diphoton decay modes, would readily show signatures of
exotic production modes: heavy particles cascade decaying to Higgses
would show up as an edge feature in the $p_T$ distribution of Higgs
candidates, for example.

\begin{figure}[htb]
\begin{center}
\psfrag{xaxislabel1234567890}{MET (GeV)}
\psfrag{yaxislabel1234567890}{Normalized yield}
\psfrag{topaxislabel123456789012345678901234567890}{Normalized MET
  distributions} 
\includegraphics[scale=1.0, angle=0]{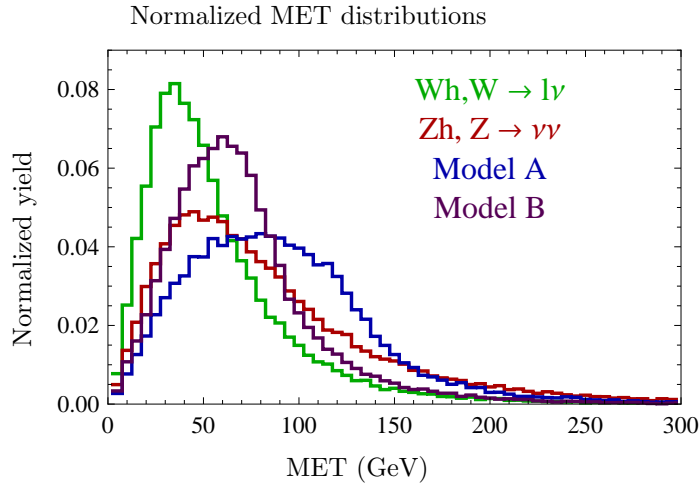}
\caption{\label{fig:MET} Truth-level MET distributions in the ATLAS
  analysis of the $\gamma \gamma$ channel for SM Higgs production via
  $Wh$, $W \to \ell \nu$ (green) and $Zh$, $Z \to \nu \nu$ (red),
  as well as exotic production benchmarks Model A (blue) and B
  (purple).}
\end{center}
\end{figure}

Using rates, however, we see that the most striking difference between
our exotic production mode and the SM production modes lies in the MET
bin efficiency.  Comparing to~\tablesref{ATLASgaga}{CMSgaga}, the only
significant contribution to the MET category in the SM comes from $Wh$
or $Zh$ production with $W \to \ell \nu$ or $Z \to \nu \nu$.  These
branching fractions suppress the MET categorization for such
production modes by about 1/3 or 1/5.  On the other hand, every event
from $\chi_1^\pm \chi_2^0$ production gives rise to two escaping LSPs,
and thus no invisible branching fraction suppression arises.  Even
with these large MET efficiencies, the benchmarks are, however,
allowed from current data, since both the ATLAS $\gamma \gamma$
analysis and the CMS mass-fit multivariate analysis have larger MET
counts compared to expectation.  For instance, ATLAS intriguingly
identifies 8 MET significance-tagged events in data compared to 4
expected from background and 1.2 expected from SM
signal~\cite{ATLAS:2013oma}.  Apart from rate, however, the MET
distributions of our new physics benchmarks are very similar to the
$Wh$, $W \to \ell \nu$ and $Zh$, $Z \to \nu \nu$ distributions, as
shown in~\figref{MET}.  Disentangling whether the MET associated with
the Higgs has non-SM contributions will be difficult unless the MET
distribution is strikingly different or we have high statistics.
Nevertheless, the possibility that Higgs bosons are produced in
association with dark matter candidates lends urgency to the need for
the experiments to publish the MET distributions of their Higgs
events.

%%%%%%%%%%%%%%%%%%%%%%%%%%%%%
%%%%%%%%%%%%%%%%%%%%%%%%%%%%%

Since the diphoton analyses from both collaborations operate as
inclusively as possible, most events that satisfy the Higgs candidate
selection end up tagged, with ``Untagged'' being the catch-all
category.  This is notable because the SM gluon fusion production rate
has the largest theory uncertainty (see~\tableref{prod_xsecs},
although recent progress has been made at improving the fixed order
calculation~\cite{Anastasiou:2014vaa}).  Hence, adopting a pessimistic
perspective, any new physics exotic production that fails the checks
for leptons, MET, and high and low two-jet mass windows will readily
contaminate the untagged category, which may not register as
significant given gluon fusion uncertainties.  Nevertheless, further
kinematic checks on $p_T$ and $\eta$ of the Higgs candidate could show
striking differences from SM expectations, which is again motivation
to publish differential distributions for Higgs signal data.

The simplest perturbation to realize a model along these lines would
be to allow for $R$-parity violation in our benchmarks, whereby the
two LSPs could then decay promptly into three jets.  Then the MET
efficiency would likely drop to a negligible percentage and would
instead flow down into the untagged category.  The prospects for
disentangling such an $R$-parity violating exotic production mode for
the Higgs are dim unless further kinematic information is made
available.  We again emphasize that the Higgs $p_T$ and MET and jet
and lepton kinematics and multiplicities are critical distributions to
publish and will serve well in distinguishing exotic from SM
production modes.

Finally, we could proceed with a rate analysis based on the
efficiencies in~\tablesref{cutflowATLAS}{cutflowCMS} and the extracted
SM signal efficiencies in~\secref{expeffs} following the observed
number of events in each analysis and using~\eqnref{rates}.  This
style of rate analysis, which was similarly noted
in~\cite{Djouadi:2012rh}, would detail the possible flat directions in
trading one production mode for another.  We want to go further,
however, by convolving the Higgs couplings into the fit with the
exotic production signal efficiencies.  This is reserved for future
work.

%%%%%%%%%%%%%%%%%%%%%%%%%%%%%
%%%%%%%%%%%%%%%%%%%%%%%%%%%%%
\section{Conclusion}
\label{sec:conclusion}

We conclude by highlighting the main points of our analysis.  We
explored the phenomenology of exotic production of the SM-like Higgs.
We find that new production modes for the Higgs are motivated from new
physics models but are currently untested by the experimental rate
results.  Coupling measurements of the Higgs are currently interpreted
in limited scenarios that exclude the possibility of new production
modes.  Since the kinematics and efficiencies of exotic production
cannot be captured by a naive rescaling of the signal strength of SM
production modes, such exotic production effects are not captured in
the current Higgs coupling measurements.  Moreover, using signal
strengths of SM production modes to characterize the Higgs signal does
not capture the full breadth of new phenomena possibly present in
Higgs data, since SM signal efficiencies are held fixed.  In
particular, even the multidimensional coupling space explored in
current Higgs fits cannot accommodate an exotic production mechanism.
For example, if an excess of MET-tagged events were observed without
observing corresponding dileptonic $Z$ or a semileptonic $W$ excesses,
then the signal cannot be accommodated with simple variations of the
$Z$ or $W$ coupling to the Higgs.

We have detailed the need for dedicated probes of exotic production.
The narrow intrinsic Higgs width, along with the new CMS bound on the
Higgs width~\cite{CMS:2014ala}, and the SM expectation that it is a
pure scalar, afford us to factorize production from decay.  Searches
for exotic decays typically require many new analyses, each targetting
a new final state.  On the other hand, probes of exotic production can
be done with minimal modification to current analyses, especially
those with well-identified Higgs candidates.  In this vein, we
highlighted the $\gamma \gamma$, $ZZ^*$, and $WW^*$ analyses as
particularly relevant for further experimental study to detail the
kinematics and multiplicities of objects produced in association with
the Higgs, as well as kinematics of the Higgs candidate itself.  In
particular, MET distributions, $p_T$ and $\eta$ distributions are
highly relevant for disentangling amongst possible competing
production modes.

We have focused on chargino-neutralino cascade decays in the MSSM as
benchmark examples of exotic production.  By adjusting the mass
splitting between the heavy parent particles and the LSP close to the
Higgs mass, we have shown the complementary nature of the Higgs
discovery analyses and dedicated SUSY searches in probing this
simplified model space.  From rate information alone, probing these
benchmarks is difficult, relying only on the limited statistical power
of MET-tagged Higgs events.  Furthermore, allowing the LSPs to decay
via $R$-parity violating couplings would drastically dilute the MET
efficiency, making this exotic production mode very difficult to
distinguish from SM $Wh$ production.  This motivates the need to go
beyond rate information to explore exotic Higgs production and lends
urgency to publish differential distributions in high resolution Higgs
final state analyses.  In particular, the $R$-parity violating version
of Models A and B would be expected to introduce a flat direction in a
combined fit to SM $Wh$ and exotic production rates.  Constructing
models that mimic other SM production modes and hence introduce
further flat directions is reserved for future work.  Again, lifting
these flat directions requires going beyond fitting for rates and
instead fitting shapes of differential distributions.

In the end, we have broadened the range of possibilities for the next
set of studies of Higgs data, focusing on how to construct and probe
new production modes of the Higgs.  Future work will focus on
exploring how Higgs coupling fits should be expanded to include the
possibility of exotic production modes.  We will also further discuss
the role of differential distributions in distinguishing various model
classes of exotic production.

%%%%%%%%%%%%%%%%%%%%%%%%%%%%%
%%%%%%%%%%%%%%%%%%%%%%%%%%%%%
\section*{Acknowledgements}
\label{sec:acknowledgements}
The author is grateful to Prateek Agrawal, Wolfgang Altmannshofer,
Bogdan Dobrescu, Paddy Fox, Claudia Frugiuele, Roni Harnik, and Joe
Lykken for useful discussions.  Fermilab is operated by Fermi Research
Alliance, LLC under Contract No. De-AC02-07CH11359 with the United
States Department of Energy.

%%%%%%%%%%%%%%%%%%%%%%%%%%%%%
%%%%%%%%%%%%%%%%%%%%%%%%%%%%%
\begin{appendix}

\section{Extrapolated efficiencies}
\label{sec:expeffs}

In this Appendix, we provide extrapolated experimental efficiencies
based on the expected Higgs signal sensitivity of each ATLAS and CMS
$\gamma \gamma$, $ZZ^* \to 4 \ell$, $WW^* \to \ell \nu \ell \nu$
analysis.  Exact references are provided in each table caption.

\begin{table}[tbh]
\renewcommand{\arraystretch}{1.2}
\begin{tabular}{|c|c|c|c|c|c|}
\hline
                 & ggF & VBF & WH & ZH & $t\bar{t}H$ \\
$N_\text{events}$ for 20.7 fb$^{-1}$   
                        & 888.2  & 73.5   & 31.9   & 18.9   & 5.9    \\ 
\hline
Lepton                  & 0.0\%  & 0.0\%  & 5.3\%  & 2.2\%  & 8.5\%  \\
$E_T$ miss significance & 0.0\%  & 0.0\%  & 1.3\%  & 3.0\%  & 2.4\%  \\
Low-mass two-jet        & 0.2\%  & 0.0\%  & 2.9\%  & 2.9\%  & 0.0\%  \\
Tight high-mass two-jet & 0.2\%  & 8.0\%  & 0.0\%  & 0.0\%  & 0.0\%  \\
Loose high-mass two-jet & 0.3\%  & 3.7\%  & 0.0\%  & 0.0\%  & 0.0\%  \\
Untagged                & 36.0\% & 25.8\% & 21.9\% & 22.2\% & 17.0\% \\
%All categories          & 36.7\% & 37.5\% & 31.5\% & 30.3\% & 27.9\% \\
\hline
\end{tabular}
\caption{Efficiency table for the ATLAS $\gamma \gamma$ analysis
  derived from Table 5 of Ref.~\cite{Aad:2013wqa}.  The mass window,
  centered at 126.5 GeV, is expected to contain 90\% of the signal.}
\label{table:ATLASgaga}
\end{table}

\begin{table}[tbh]
\renewcommand{\arraystretch}{1.2}
\begin{tabular}{|c|c|c|c|c|c|}
\hline
                 & ggF+$t\bar{t}H$ & VBF & WH+ZH \\
$N_\text{events}$ for 20.7 fb$^{-1}$   
                        & 50.8   & 4.1    & 2.8    \\
\hline
ggF-like                & 26.6\% & 19.3\% & 23.0\% \\
VBF-like                & 0.6\%  & 10.5\% & 0.4\%  \\
VH-like                 & 0.1\%  & 0.0\%  & 5.0\%  \\
%All categories          & 27.2\% & 29.8\% & 28.4\% \\
\hline
\end{tabular}
\caption{Efficiency table for the ATLAS $Z Z^*$ analysis derived from
  Table 2 of Ref.~\cite{ATLAS:2013nma}, with $m_H = 125$ GeV.}
\label{table:ATLASZZ}
\end{table}

\begin{table}[tbh]
\renewcommand{\arraystretch}{1.2}
\begin{tabular}{|c|c|c|c|c|c|}
\hline
                 & Signal \\ %ggF+$t\bar{t}H$ & VBF & WH+ZH \\
$N_\text{events}$ for 20.7 fb$^{-1}$   
                     & 11029 \\
\hline
0-jet                & 0.907\%   \\
1-jet                & 0.372\%  \\
$\geq$ 2-jet         & 0.099\%  \\
%All categories       & 1.377\% \\
\hline
\end{tabular}
\caption{Efficiency table for the ATLAS $W W^*$ analysis derived from
  Table 9 of Ref.~\cite{Aad:2013wqa}, with $m_H = 125.5$ GeV.  Event
  counts are given in the transverse mass region $0.75 m_H < m_T <
  m_H$ for $N_{\text{jet}} \leq 1$ and $m_T < 1.2 m_H$ for
  $N_{\text{jet}} >= 2$.  Note here that we normalized the $h \to WW^*
  \to \ell \nu \ell \nu$ rate to include $\ell = e$, $\mu$, $\tau$.}
\label{table:ATLASWW}
\end{table}

\begin{table}[tbh]
\renewcommand{\arraystretch}{1.2}
\begin{tabular}{|c|c|c|c|c|c|}
\hline
                        & ggF    & VBF    & VH     & $t\bar{t}H$ \\
$N_\text{events}$ for 19.6 fb$^{-1}$   
                        & 861.1  & 70.5   & 50.0   & 5.8    \\ 
\hline
Muon                    & 0.0\%  & $< 0.1$\%  & 2.2\%  & 5.0\% \\
Electron                & $< 0.1$\%  & $< 0.1$\%  & 1.4\%  & 3.1\% \\
Dijet tight             & 0.2\%  & 10.3\% & $< 0.1$\%  & 0.2\%  \\
Dijet loose             & 0.6\%  & 8.3\%  & 0.4\%  & 1.0\%  \\
$E_T$ miss              & $< 0.1$\%  & $< 0.1$\%  & 2.2\%  & 3.4\% \\
Untagged combined       & 38.3\% & 25.7\% & 31.1\% & 32.9\% \\
%All categories          & 38.3\% & 25.7\% & 31.1\% & 32.9\% \\
\hline
\end{tabular}
\caption{Efficiency table for the CMS $\gamma \gamma$ analysis derived
  from Table 5 of Ref.~\cite{CMS:ril}, which details the mass-fit MVA
  analysis, assuming a signal $m_H = 125$ GeV.}
\label{table:CMSgaga}
\end{table}

\begin{table}[tbh]
\renewcommand{\arraystretch}{1.2}
\begin{tabular}{|c|c|c|c|c|c|}
\hline
                        & ggF    & VBF    & WH     & ZH     & $t\bar{t}H$ \\
$N_\text{events}$ for 5.1 fb$^{-1}$ + 19.6 fb$^{-1}$   
                        & 60.9   & 5.0    & 2.2    & 1.3    & 0.4    \\ 
\hline
0/1-jet                 & 25.3\% & 14.0\% & 12.6\% & 16.1\% & 0.0\%  \\
Dijet                   & 2.6\%  & 17.3\% & 9.5\%  & 12.3\% & 20.3\% \\
%All categories          & 27.9\% & 31.3\% & 22.1\% & 28.3\% & 20.3\% \\
\hline
\end{tabular}
\caption{Efficiency table for the CMS $Z Z^*$ analysis derived from
  Table 5 of Ref.~\cite{Chatrchyan:2013mxa}.  Results are integrated
  over the mass range 121.5 to 130.5 GeV and combine 7 and 8 TeV
  data.  The signal uses $m_H = 126$ GeV.}
\label{table:CMSZZ}
\end{table}

\begin{table}[tbh]
\renewcommand{\arraystretch}{1.2}
\begin{tabular}{|c|c|c|c|c|c|}
\hline
                        & ggF    & VBF+VH \\
$N_\text{events}$ for 19.4 fb$^{-1}$   
                        & 8852   & 1212  \\
\hline
0-jet                   & 1.62\% & 0.27\% \\
1-jet                   & 0.60\% & 0.79\% \\
%All categories          & 2.21\% & 1.07\% \\
\hline
\end{tabular}
\caption{Efficiency table for the CMS $W W^*$ analysis derived from
  Table 5 of Ref.~\cite{Chatrchyan:2013iaa}, adopting the signal $m_H
  = 125$ GeV row.  Note here that we normalized the $h \to \ell \nu
  \ell \nu$ rate to include $\ell = e$, $\mu$, $\tau$.}
\label{table:CMSWW}
\end{table}

\end{appendix}

\newpage

\quad

\newpage

\quad

\newpage

%%%%%%%%%%%%%%%%%%%%%%%%%%%%%
% Bibliography
%%%%%%%%%%%%%%%%%%%%%%%%%%%%%
%\bibliographystyle{plain}
%\bibliographystyle{abbrv}
\bibliographystyle{apsrev4-1}
\bibliography{Higgs}

\begin{thebibliography}{130}%
\makeatletter
\providecommand \@ifxundefined [1]{%
 \@ifx{#1\undefined}
}%
\providecommand \@ifnum [1]{%
 \ifnum #1\expandafter \@firstoftwo
 \else \expandafter \@secondoftwo
 \fi
}%
\providecommand \@ifx [1]{%
 \ifx #1\expandafter \@firstoftwo
 \else \expandafter \@secondoftwo
 \fi
}%
\providecommand \natexlab [1]{#1}%
\providecommand \enquote  [1]{``#1''}%
\providecommand \bibnamefont  [1]{#1}%
\providecommand \bibfnamefont [1]{#1}%
\providecommand \citenamefont [1]{#1}%
\providecommand \href@noop [0]{\@secondoftwo}%
\providecommand \href [0]{\begingroup \@sanitize@url \@href}%
\providecommand \@href[1]{\@@startlink{#1}\@@href}%
\providecommand \@@href[1]{\endgroup#1\@@endlink}%
\providecommand \@sanitize@url [0]{\catcode `\\12\catcode `\$12\catcode
  `\&12\catcode `\#12\catcode `\^12\catcode `\_12\catcode `\%12\relax}%
\providecommand \@@startlink[1]{}%
\providecommand \@@endlink[0]{}%
\providecommand \url  [0]{\begingroup\@sanitize@url \@url }%
\providecommand \@url [1]{\endgroup\@href {#1}{\urlprefix }}%
\providecommand \urlprefix  [0]{URL }%
\providecommand \Eprint [0]{\href }%
\providecommand \doibase [0]{http://dx.doi.org/}%
\providecommand \selectlanguage [0]{\@gobble}%
\providecommand \bibinfo  [0]{\@secondoftwo}%
\providecommand \bibfield  [0]{\@secondoftwo}%
\providecommand \translation [1]{[#1]}%
\providecommand \BibitemOpen [0]{}%
\providecommand \bibitemStop [0]{}%
\providecommand \bibitemNoStop [0]{.\EOS\space}%
\providecommand \EOS [0]{\spacefactor3000\relax}%
\providecommand \BibitemShut  [1]{\csname bibitem#1\endcsname}%
\let\auto@bib@innerbib\@empty
%</preamble>
\bibitem [{\citenamefont {Chatrchyan}\ \emph
  {et~al.}(2013{\natexlab{a}})\citenamefont {Chatrchyan} \emph
  {et~al.}}]{CMS:ril}%
  \BibitemOpen
  \bibfield  {author} {\bibinfo {author} {\bibfnamefont {S.}~\bibnamefont
  {Chatrchyan}} \emph {et~al.} (\bibinfo {collaboration} {CMS Collaboration}),\
  }\href@noop {} {\  (\bibinfo {year} {2013}{\natexlab{a}})},\ \bibinfo {note}
  {{C}MS-PAS-HIG-13-001}\BibitemShut {NoStop}%
%%CITATION = CMS-PAS-HIG-13-001 ETC.;%%
\bibitem [{\citenamefont {Chatrchyan}\ \emph
  {et~al.}(2013{\natexlab{b}})\citenamefont {Chatrchyan} \emph
  {et~al.}}]{Chatrchyan:2013mxa}%
  \BibitemOpen
  \bibfield  {author} {\bibinfo {author} {\bibfnamefont {S.}~\bibnamefont
  {Chatrchyan}} \emph {et~al.} (\bibinfo {collaboration} {CMS Collaboration}),\
  }\href@noop {} {\  (\bibinfo {year} {2013}{\natexlab{b}})},\ \Eprint
  {http://arxiv.org/abs/1312.5353} {arXiv:1312.5353 [hep-ex]} \BibitemShut
  {NoStop}%
%%CITATION = ARXIV:1312.5353;%%
\bibitem [{\citenamefont {Chatrchyan}\ \emph
  {et~al.}(2013{\natexlab{c}})\citenamefont {Chatrchyan} \emph
  {et~al.}}]{CMS:bxa}%
  \BibitemOpen
  \bibfield  {author} {\bibinfo {author} {\bibfnamefont {S.}~\bibnamefont
  {Chatrchyan}} \emph {et~al.} (\bibinfo {collaboration} {CMS Collaboration}),\
  }\href@noop {} {\  (\bibinfo {year} {2013}{\natexlab{c}})},\ \bibinfo {note}
  {{C}MS-PAS-HIG-13-003}\BibitemShut {NoStop}%
%%CITATION = CMS-PAS-HIG-13-003 ETC.;%%
\bibitem [{\citenamefont {Aad}\ \emph {et~al.}(2013{\natexlab{a}})\citenamefont
  {Aad} \emph {et~al.}}]{ATLAS:2013oma}%
  \BibitemOpen
  \bibfield  {author} {\bibinfo {author} {\bibfnamefont {G.}~\bibnamefont
  {Aad}} \emph {et~al.} (\bibinfo {collaboration} {ATLAS Collaboration}),\
  }\href@noop {} {\  (\bibinfo {year} {2013}{\natexlab{a}})},\ \bibinfo {note}
  {{A}TLAS-CONF-2013-012}\BibitemShut {NoStop}%
%%CITATION = ATLAS-CONF-2013-012 ETC.;%%
\bibitem [{\citenamefont {Aad}\ \emph {et~al.}(2013{\natexlab{b}})\citenamefont
  {Aad} \emph {et~al.}}]{ATLAS:2013nma}%
  \BibitemOpen
  \bibfield  {author} {\bibinfo {author} {\bibfnamefont {G.}~\bibnamefont
  {Aad}} \emph {et~al.} (\bibinfo {collaboration} {ATLAS Collaboration}),\
  }\href@noop {} {\  (\bibinfo {year} {2013}{\natexlab{b}})},\ \bibinfo {note}
  {{A}TLAS-CONF-2013-013}\BibitemShut {NoStop}%
%%CITATION = ATLAS-CONF-2013-013 ETC.;%%
\bibitem [{\citenamefont {Aad}\ \emph {et~al.}(2013{\natexlab{c}})\citenamefont
  {Aad} \emph {et~al.}}]{ATLAS:2013wla}%
  \BibitemOpen
  \bibfield  {author} {\bibinfo {author} {\bibfnamefont {G.}~\bibnamefont
  {Aad}} \emph {et~al.} (\bibinfo {collaboration} {ATLAS Collaboration}),\
  }\href@noop {} {\  (\bibinfo {year} {2013}{\natexlab{c}})},\ \bibinfo {note}
  {{A}TLAS-CONF-2013-030}\BibitemShut {NoStop}%
%%CITATION = ATLAS-CONF-2013-030 ETC.;%%
\bibitem [{\citenamefont {Dawson}\ \emph {et~al.}(2013)\citenamefont {Dawson},
  \citenamefont {Gritsan}, \citenamefont {Logan}, \citenamefont {Qian},
  \citenamefont {Tully} \emph {et~al.}}]{Dawson:2013bba}%
  \BibitemOpen
  \bibfield  {author} {\bibinfo {author} {\bibfnamefont {S.}~\bibnamefont
  {Dawson}}, \bibinfo {author} {\bibfnamefont {A.}~\bibnamefont {Gritsan}},
  \bibinfo {author} {\bibfnamefont {H.}~\bibnamefont {Logan}}, \bibinfo
  {author} {\bibfnamefont {J.}~\bibnamefont {Qian}}, \bibinfo {author}
  {\bibfnamefont {C.}~\bibnamefont {Tully}},  \emph {et~al.},\ }\href@noop {}
  {\  (\bibinfo {year} {2013})},\ \Eprint {http://arxiv.org/abs/1310.8361}
  {arXiv:1310.8361 [hep-ex]} \BibitemShut {NoStop}%
%%CITATION = ARXIV:1310.8361;%%
\bibitem [{\citenamefont {Azatov}\ \emph
  {et~al.}(2012{\natexlab{a}})\citenamefont {Azatov}, \citenamefont {Contino},\
  and\ \citenamefont {Galloway}}]{Azatov:2012bz}%
  \BibitemOpen
  \bibfield  {author} {\bibinfo {author} {\bibfnamefont {A.}~\bibnamefont
  {Azatov}}, \bibinfo {author} {\bibfnamefont {R.}~\bibnamefont {Contino}}, \
  and\ \bibinfo {author} {\bibfnamefont {J.}~\bibnamefont {Galloway}},\ }\href
  {\doibase 10.1007/JHEP04(2012)127, 10.1007/JHEP04(2013)140} {\bibfield
  {journal} {\bibinfo  {journal} {JHEP}\ }\textbf {\bibinfo {volume} {1204}},\
  \bibinfo {pages} {127} (\bibinfo {year} {2012}{\natexlab{a}})},\ \Eprint
  {http://arxiv.org/abs/1202.3415} {arXiv:1202.3415 [hep-ph]} \BibitemShut
  {NoStop}%
%%CITATION = ARXIV:1202.3415;%%
\bibitem [{\citenamefont {Espinosa}\ \emph {et~al.}(2012)\citenamefont
  {Espinosa}, \citenamefont {Grojean}, \citenamefont {Muhlleitner},\ and\
  \citenamefont {Trott}}]{Espinosa:2012ir}%
  \BibitemOpen
  \bibfield  {author} {\bibinfo {author} {\bibfnamefont {J.}~\bibnamefont
  {Espinosa}}, \bibinfo {author} {\bibfnamefont {C.}~\bibnamefont {Grojean}},
  \bibinfo {author} {\bibfnamefont {M.}~\bibnamefont {Muhlleitner}}, \ and\
  \bibinfo {author} {\bibfnamefont {M.}~\bibnamefont {Trott}},\ }\href
  {\doibase 10.1007/JHEP05(2012)097} {\bibfield  {journal} {\bibinfo  {journal}
  {JHEP}\ }\textbf {\bibinfo {volume} {1205}},\ \bibinfo {pages} {097}
  (\bibinfo {year} {2012})},\ \Eprint {http://arxiv.org/abs/1202.3697}
  {arXiv:1202.3697 [hep-ph]} \BibitemShut {NoStop}%
%%CITATION = ARXIV:1202.3697;%%
\bibitem [{\citenamefont {Azatov}\ \emph
  {et~al.}(2012{\natexlab{b}})\citenamefont {Azatov}, \citenamefont {Contino},
  \citenamefont {Del~Re}, \citenamefont {Galloway}, \citenamefont {Grassi}
  \emph {et~al.}}]{Azatov:2012rd}%
  \BibitemOpen
  \bibfield  {author} {\bibinfo {author} {\bibfnamefont {A.}~\bibnamefont
  {Azatov}}, \bibinfo {author} {\bibfnamefont {R.}~\bibnamefont {Contino}},
  \bibinfo {author} {\bibfnamefont {D.}~\bibnamefont {Del~Re}}, \bibinfo
  {author} {\bibfnamefont {J.}~\bibnamefont {Galloway}}, \bibinfo {author}
  {\bibfnamefont {M.}~\bibnamefont {Grassi}},  \emph {et~al.},\ }\href
  {\doibase 10.1007/JHEP06(2012)134} {\bibfield  {journal} {\bibinfo  {journal}
  {JHEP}\ }\textbf {\bibinfo {volume} {1206}},\ \bibinfo {pages} {134}
  (\bibinfo {year} {2012}{\natexlab{b}})},\ \Eprint
  {http://arxiv.org/abs/1204.4817} {arXiv:1204.4817 [hep-ph]} \BibitemShut
  {NoStop}%
%%CITATION = ARXIV:1204.4817;%%
\bibitem [{\citenamefont {Klute}\ \emph {et~al.}(2012)\citenamefont {Klute},
  \citenamefont {Lafaye}, \citenamefont {Plehn}, \citenamefont {Rauch},\ and\
  \citenamefont {Zerwas}}]{Klute:2012pu}%
  \BibitemOpen
  \bibfield  {author} {\bibinfo {author} {\bibfnamefont {M.}~\bibnamefont
  {Klute}}, \bibinfo {author} {\bibfnamefont {R.}~\bibnamefont {Lafaye}},
  \bibinfo {author} {\bibfnamefont {T.}~\bibnamefont {Plehn}}, \bibinfo
  {author} {\bibfnamefont {M.}~\bibnamefont {Rauch}}, \ and\ \bibinfo {author}
  {\bibfnamefont {D.}~\bibnamefont {Zerwas}},\ }\href {\doibase
  10.1103/PhysRevLett.109.101801} {\bibfield  {journal} {\bibinfo  {journal}
  {Phys.Rev.Lett.}\ }\textbf {\bibinfo {volume} {109}},\ \bibinfo {pages}
  {101801} (\bibinfo {year} {2012})},\ \Eprint {http://arxiv.org/abs/1205.2699}
  {arXiv:1205.2699 [hep-ph]} \BibitemShut {NoStop}%
%%CITATION = ARXIV:1205.2699;%%
\bibitem [{\citenamefont {Carmi}\ \emph {et~al.}(2013)\citenamefont {Carmi},
  \citenamefont {Falkowski}, \citenamefont {Kuflik},\ and\ \citenamefont
  {Volansky}}]{Carmi:2012zd}%
  \BibitemOpen
  \bibfield  {author} {\bibinfo {author} {\bibfnamefont {D.}~\bibnamefont
  {Carmi}}, \bibinfo {author} {\bibfnamefont {A.}~\bibnamefont {Falkowski}},
  \bibinfo {author} {\bibfnamefont {E.}~\bibnamefont {Kuflik}}, \ and\ \bibinfo
  {author} {\bibfnamefont {T.}~\bibnamefont {Volansky}},\ }\href@noop {}
  {\bibfield  {journal} {\bibinfo  {journal} {Frascati Phys.Ser.}\ }\textbf
  {\bibinfo {volume} {57}},\ \bibinfo {pages} {315} (\bibinfo {year} {2013})},\
  \Eprint {http://arxiv.org/abs/1206.4201} {arXiv:1206.4201 [hep-ph]}
  \BibitemShut {NoStop}%
%%CITATION = ARXIV:1206.4201;%%
\bibitem [{\citenamefont {Blum}\ \emph {et~al.}(2013)\citenamefont {Blum},
  \citenamefont {D'Agnolo},\ and\ \citenamefont {Fan}}]{Blum:2012ii}%
  \BibitemOpen
  \bibfield  {author} {\bibinfo {author} {\bibfnamefont {K.}~\bibnamefont
  {Blum}}, \bibinfo {author} {\bibfnamefont {R.~T.}\ \bibnamefont {D'Agnolo}},
  \ and\ \bibinfo {author} {\bibfnamefont {J.}~\bibnamefont {Fan}},\ }\href
  {\doibase 10.1007/JHEP01(2013)057} {\bibfield  {journal} {\bibinfo  {journal}
  {JHEP}\ }\textbf {\bibinfo {volume} {1301}},\ \bibinfo {pages} {057}
  (\bibinfo {year} {2013})},\ \Eprint {http://arxiv.org/abs/1206.5303}
  {arXiv:1206.5303 [hep-ph]} \BibitemShut {NoStop}%
%%CITATION = ARXIV:1206.5303;%%
\bibitem [{\citenamefont {Dobrescu}\ and\ \citenamefont
  {Lykken}(2013)}]{Dobrescu:2012td}%
  \BibitemOpen
  \bibfield  {author} {\bibinfo {author} {\bibfnamefont {B.~A.}\ \bibnamefont
  {Dobrescu}}\ and\ \bibinfo {author} {\bibfnamefont {J.~D.}\ \bibnamefont
  {Lykken}},\ }\href {\doibase 10.1007/JHEP02(2013)073} {\bibfield  {journal}
  {\bibinfo  {journal} {JHEP}\ }\textbf {\bibinfo {volume} {1302}},\ \bibinfo
  {pages} {073} (\bibinfo {year} {2013})},\ \Eprint
  {http://arxiv.org/abs/1210.3342} {arXiv:1210.3342 [hep-ph]} \BibitemShut
  {NoStop}%
%%CITATION = ARXIV:1210.3342;%%
\bibitem [{\citenamefont {Carmi}\ \emph {et~al.}(2012)\citenamefont {Carmi},
  \citenamefont {Falkowski}, \citenamefont {Kuflik}, \citenamefont {Volansky},\
  and\ \citenamefont {Zupan}}]{Carmi:2012in}%
  \BibitemOpen
  \bibfield  {author} {\bibinfo {author} {\bibfnamefont {D.}~\bibnamefont
  {Carmi}}, \bibinfo {author} {\bibfnamefont {A.}~\bibnamefont {Falkowski}},
  \bibinfo {author} {\bibfnamefont {E.}~\bibnamefont {Kuflik}}, \bibinfo
  {author} {\bibfnamefont {T.}~\bibnamefont {Volansky}}, \ and\ \bibinfo
  {author} {\bibfnamefont {J.}~\bibnamefont {Zupan}},\ }\href {\doibase
  10.1007/JHEP10(2012)196} {\bibfield  {journal} {\bibinfo  {journal} {JHEP}\
  }\textbf {\bibinfo {volume} {1210}},\ \bibinfo {pages} {196} (\bibinfo {year}
  {2012})},\ \Eprint {http://arxiv.org/abs/1207.1718} {arXiv:1207.1718
  [hep-ph]} \BibitemShut {NoStop}%
%%CITATION = ARXIV:1207.1718;%%
\bibitem [{\citenamefont {Plehn}\ and\ \citenamefont
  {Rauch}(2012)}]{Plehn:2012iz}%
  \BibitemOpen
  \bibfield  {author} {\bibinfo {author} {\bibfnamefont {T.}~\bibnamefont
  {Plehn}}\ and\ \bibinfo {author} {\bibfnamefont {M.}~\bibnamefont {Rauch}},\
  }\href {\doibase 10.1209/0295-5075/100/11002} {\bibfield  {journal} {\bibinfo
   {journal} {Europhys.Lett.}\ }\textbf {\bibinfo {volume} {100}},\ \bibinfo
  {pages} {11002} (\bibinfo {year} {2012})},\ \Eprint
  {http://arxiv.org/abs/1207.6108} {arXiv:1207.6108 [hep-ph]} \BibitemShut
  {NoStop}%
%%CITATION = ARXIV:1207.6108;%%
\bibitem [{\citenamefont {David}\ \emph {et~al.}(2012)\citenamefont {David}
  \emph {et~al.}}]{LHCHiggsCrossSectionWorkingGroup:2012nn}%
  \BibitemOpen
  \bibfield  {author} {\bibinfo {author} {\bibfnamefont {A.}~\bibnamefont
  {David}} \emph {et~al.} (\bibinfo {collaboration} {LHC Higgs Cross Section
  Working Group}),\ }\href@noop {} {\  (\bibinfo {year} {2012})},\ \Eprint
  {http://arxiv.org/abs/1209.0040} {arXiv:1209.0040 [hep-ph]} \BibitemShut
  {NoStop}%
%%CITATION = ARXIV:1209.0040;%%
\bibitem [{\citenamefont {Corbett}\ \emph
  {et~al.}(2013{\natexlab{a}})\citenamefont {Corbett}, \citenamefont {Eboli},
  \citenamefont {Gonzalez-Fraile},\ and\ \citenamefont
  {Gonzalez-Garcia}}]{Corbett:2012ja}%
  \BibitemOpen
  \bibfield  {author} {\bibinfo {author} {\bibfnamefont {T.}~\bibnamefont
  {Corbett}}, \bibinfo {author} {\bibfnamefont {O.}~\bibnamefont {Eboli}},
  \bibinfo {author} {\bibfnamefont {J.}~\bibnamefont {Gonzalez-Fraile}}, \ and\
  \bibinfo {author} {\bibfnamefont {M.}~\bibnamefont {Gonzalez-Garcia}},\
  }\href {\doibase 10.1103/PhysRevD.87.015022} {\bibfield  {journal} {\bibinfo
  {journal} {Phys.Rev.}\ }\textbf {\bibinfo {volume} {D87}},\ \bibinfo {pages}
  {015022} (\bibinfo {year} {2013}{\natexlab{a}})},\ \Eprint
  {http://arxiv.org/abs/1211.4580} {arXiv:1211.4580 [hep-ph]} \BibitemShut
  {NoStop}%
%%CITATION = ARXIV:1211.4580;%%
\bibitem [{\citenamefont {Belanger}\ \emph {et~al.}(2013)\citenamefont
  {Belanger}, \citenamefont {Dumont}, \citenamefont {Ellwanger}, \citenamefont
  {Gunion},\ and\ \citenamefont {Kraml}}]{Belanger:2013kya}%
  \BibitemOpen
  \bibfield  {author} {\bibinfo {author} {\bibfnamefont {G.}~\bibnamefont
  {Belanger}}, \bibinfo {author} {\bibfnamefont {B.}~\bibnamefont {Dumont}},
  \bibinfo {author} {\bibfnamefont {U.}~\bibnamefont {Ellwanger}}, \bibinfo
  {author} {\bibfnamefont {J.}~\bibnamefont {Gunion}}, \ and\ \bibinfo {author}
  {\bibfnamefont {S.}~\bibnamefont {Kraml}},\ }\href {\doibase
  10.1016/j.physletb.2013.05.024} {\bibfield  {journal} {\bibinfo  {journal}
  {Phys.Lett.}\ }\textbf {\bibinfo {volume} {B723}},\ \bibinfo {pages} {340}
  (\bibinfo {year} {2013})},\ \Eprint {http://arxiv.org/abs/1302.5694}
  {arXiv:1302.5694 [hep-ph]} \BibitemShut {NoStop}%
%%CITATION = ARXIV:1302.5694;%%
\bibitem [{\citenamefont {Falkowski}\ \emph {et~al.}(2013)\citenamefont
  {Falkowski}, \citenamefont {Riva},\ and\ \citenamefont
  {Urbano}}]{Falkowski:2013dza}%
  \BibitemOpen
  \bibfield  {author} {\bibinfo {author} {\bibfnamefont {A.}~\bibnamefont
  {Falkowski}}, \bibinfo {author} {\bibfnamefont {F.}~\bibnamefont {Riva}}, \
  and\ \bibinfo {author} {\bibfnamefont {A.}~\bibnamefont {Urbano}},\ }\href
  {\doibase 10.1007/JHEP11(2013)111} {\bibfield  {journal} {\bibinfo  {journal}
  {JHEP}\ }\textbf {\bibinfo {volume} {1311}},\ \bibinfo {pages} {111}
  (\bibinfo {year} {2013})},\ \Eprint {http://arxiv.org/abs/1303.1812}
  {arXiv:1303.1812 [hep-ph]} \BibitemShut {NoStop}%
%%CITATION = ARXIV:1303.1812;%%
\bibitem [{\citenamefont {Giardino}\ \emph {et~al.}(2013)\citenamefont
  {Giardino}, \citenamefont {Kannike}, \citenamefont {Masina}, \citenamefont
  {Raidal},\ and\ \citenamefont {Strumia}}]{Giardino:2013bma}%
  \BibitemOpen
  \bibfield  {author} {\bibinfo {author} {\bibfnamefont {P.~P.}\ \bibnamefont
  {Giardino}}, \bibinfo {author} {\bibfnamefont {K.}~\bibnamefont {Kannike}},
  \bibinfo {author} {\bibfnamefont {I.}~\bibnamefont {Masina}}, \bibinfo
  {author} {\bibfnamefont {M.}~\bibnamefont {Raidal}}, \ and\ \bibinfo {author}
  {\bibfnamefont {A.}~\bibnamefont {Strumia}},\ }\href@noop {} {\  (\bibinfo
  {year} {2013})},\ \Eprint {http://arxiv.org/abs/1303.3570} {arXiv:1303.3570
  [hep-ph]} \BibitemShut {NoStop}%
%%CITATION = ARXIV:1303.3570;%%
\bibitem [{\citenamefont {Djouadi}\ and\ \citenamefont
  {Moreau}(2013)}]{Djouadi:2013qya}%
  \BibitemOpen
  \bibfield  {author} {\bibinfo {author} {\bibfnamefont {A.}~\bibnamefont
  {Djouadi}}\ and\ \bibinfo {author} {\bibfnamefont {G.}~\bibnamefont
  {Moreau}},\ }\href@noop {} {\  (\bibinfo {year} {2013})},\ \Eprint
  {http://arxiv.org/abs/1303.6591} {arXiv:1303.6591 [hep-ph]} \BibitemShut
  {NoStop}%
%%CITATION = ARXIV:1303.6591;%%
\bibitem [{\citenamefont {Corbett}\ \emph
  {et~al.}(2013{\natexlab{b}})\citenamefont {Corbett}, \citenamefont {Eboli},
  \citenamefont {Gonzalez-Fraile},\ and\ \citenamefont
  {Gonzalez-Garcia}}]{Corbett:2013hia}%
  \BibitemOpen
  \bibfield  {author} {\bibinfo {author} {\bibfnamefont {T.}~\bibnamefont
  {Corbett}}, \bibinfo {author} {\bibfnamefont {O.}~\bibnamefont {Eboli}},
  \bibinfo {author} {\bibfnamefont {J.}~\bibnamefont {Gonzalez-Fraile}}, \ and\
  \bibinfo {author} {\bibfnamefont {M.}~\bibnamefont {Gonzalez-Garcia}},\
  }\href@noop {} {\  (\bibinfo {year} {2013}{\natexlab{b}})},\ \Eprint
  {http://arxiv.org/abs/1306.0006} {arXiv:1306.0006 [hep-ph]} \BibitemShut
  {NoStop}%
%%CITATION = ARXIV:1306.0006;%%
\bibitem [{\citenamefont {Artoisenet}\ \emph {et~al.}(2013)\citenamefont
  {Artoisenet}, \citenamefont {de~Aquino}, \citenamefont {Demartin},
  \citenamefont {Frederix}, \citenamefont {Frixione} \emph
  {et~al.}}]{Artoisenet:2013puc}%
  \BibitemOpen
  \bibfield  {author} {\bibinfo {author} {\bibfnamefont {P.}~\bibnamefont
  {Artoisenet}}, \bibinfo {author} {\bibfnamefont {P.}~\bibnamefont
  {de~Aquino}}, \bibinfo {author} {\bibfnamefont {F.}~\bibnamefont {Demartin}},
  \bibinfo {author} {\bibfnamefont {R.}~\bibnamefont {Frederix}}, \bibinfo
  {author} {\bibfnamefont {S.}~\bibnamefont {Frixione}},  \emph {et~al.},\
  }\href {\doibase 10.1007/JHEP11(2013)043} {\bibfield  {journal} {\bibinfo
  {journal} {JHEP}\ }\textbf {\bibinfo {volume} {1311}},\ \bibinfo {pages}
  {043} (\bibinfo {year} {2013})},\ \Eprint {http://arxiv.org/abs/1306.6464}
  {arXiv:1306.6464 [hep-ph]} \BibitemShut {NoStop}%
%%CITATION = ARXIV:1306.6464;%%
\bibitem [{\citenamefont {Pomarol}\ and\ \citenamefont
  {Riva}(2013)}]{Pomarol:2013zra}%
  \BibitemOpen
  \bibfield  {author} {\bibinfo {author} {\bibfnamefont {A.}~\bibnamefont
  {Pomarol}}\ and\ \bibinfo {author} {\bibfnamefont {F.}~\bibnamefont {Riva}},\
  }\href@noop {} {\  (\bibinfo {year} {2013})},\ \Eprint
  {http://arxiv.org/abs/1308.2803} {arXiv:1308.2803 [hep-ph]} \BibitemShut
  {NoStop}%
%%CITATION = ARXIV:1308.2803;%%
\bibitem [{\citenamefont {Boos}\ \emph {et~al.}(2013)\citenamefont {Boos},
  \citenamefont {Bunichev}, \citenamefont {Dubinin},\ and\ \citenamefont
  {Kurihara}}]{Boos:2013mqa}%
  \BibitemOpen
  \bibfield  {author} {\bibinfo {author} {\bibfnamefont {E.}~\bibnamefont
  {Boos}}, \bibinfo {author} {\bibfnamefont {V.}~\bibnamefont {Bunichev}},
  \bibinfo {author} {\bibfnamefont {M.}~\bibnamefont {Dubinin}}, \ and\
  \bibinfo {author} {\bibfnamefont {Y.}~\bibnamefont {Kurihara}},\ }\href@noop
  {} {\  (\bibinfo {year} {2013})},\ \Eprint {http://arxiv.org/abs/1309.5410}
  {arXiv:1309.5410 [hep-ph]} \BibitemShut {NoStop}%
%%CITATION = ARXIV:1309.5410;%%
\bibitem [{\citenamefont {Stål}\ and\ \citenamefont
  {Stefaniak}(2013)}]{Stal:2013hwa}%
  \BibitemOpen
  \bibfield  {author} {\bibinfo {author} {\bibfnamefont {O.}~\bibnamefont
  {Stål}}\ and\ \bibinfo {author} {\bibfnamefont {T.}~\bibnamefont
  {Stefaniak}},\ }\href@noop {} {\  (\bibinfo {year} {2013})},\ \Eprint
  {http://arxiv.org/abs/1310.4039} {arXiv:1310.4039 [hep-ph]} \BibitemShut
  {NoStop}%
%%CITATION = ARXIV:1310.4039;%%
\bibitem [{\citenamefont {Aad}\ \emph {et~al.}(2013{\natexlab{d}})\citenamefont
  {Aad} \emph {et~al.}}]{ATLAS:2013mma}%
  \BibitemOpen
  \bibfield  {author} {\bibinfo {author} {\bibfnamefont {G.}~\bibnamefont
  {Aad}} \emph {et~al.} (\bibinfo {collaboration} {ATLAS Collaboration}),\
  }\href@noop {} {\  (\bibinfo {year} {2013}{\natexlab{d}})},\ \bibinfo {note}
  {{A}TLAS-CONF-2013-014}\BibitemShut {NoStop}%
%%CITATION = ATLAS-CONF-2013-014 ETC.;%%
\bibitem [{\citenamefont {Aad}\ \emph {et~al.}(2013{\natexlab{e}})\citenamefont
  {Aad} \emph {et~al.}}]{ATLAS:2013sla}%
  \BibitemOpen
  \bibfield  {author} {\bibinfo {author} {\bibfnamefont {G.}~\bibnamefont
  {Aad}} \emph {et~al.} (\bibinfo {collaboration} {ATLAS Collaboration}),\
  }\href@noop {} {\  (\bibinfo {year} {2013}{\natexlab{e}})},\ \bibinfo {note}
  {{A}TLAS-CONF-2013-034}\BibitemShut {NoStop}%
%%CITATION = ATLAS-CONF-2013-034 ETC.;%%
\bibitem [{\citenamefont {Aad}\ \emph {et~al.}(2013{\natexlab{f}})\citenamefont
  {Aad} \emph {et~al.}}]{Aad:2013wqa}%
  \BibitemOpen
  \bibfield  {author} {\bibinfo {author} {\bibfnamefont {G.}~\bibnamefont
  {Aad}} \emph {et~al.} (\bibinfo {collaboration} {ATLAS Collaboration}),\
  }\href {\doibase 10.1016/j.physletb.2013.08.010} {\bibfield  {journal}
  {\bibinfo  {journal} {Phys.Lett.}\ }\textbf {\bibinfo {volume} {B726}},\
  \bibinfo {pages} {88} (\bibinfo {year} {2013}{\natexlab{f}})},\ \Eprint
  {http://arxiv.org/abs/1307.1427} {arXiv:1307.1427 [hep-ex]} \BibitemShut
  {NoStop}%
%%CITATION = ARXIV:1307.1427;%%
\bibitem [{\citenamefont {Chatrchyan}\ \emph
  {et~al.}(2013{\natexlab{d}})\citenamefont {Chatrchyan} \emph
  {et~al.}}]{CMS:yva}%
  \BibitemOpen
  \bibfield  {author} {\bibinfo {author} {\bibfnamefont {S.}~\bibnamefont
  {Chatrchyan}} \emph {et~al.} (\bibinfo {collaboration} {CMS Collaboration}),\
  }\href@noop {} {\  (\bibinfo {year} {2013}{\natexlab{d}})},\ \bibinfo {note}
  {{C}MS-PAS-HIG-13-005}\BibitemShut {NoStop}%
%%CITATION = CMS-PAS-HIG-13-005 ETC.;%%
\bibitem [{\citenamefont {Aad}\ \emph {et~al.}(2014{\natexlab{a}})\citenamefont
  {Aad} \emph {et~al.}}]{ATLAS-CONF-2014-010}%
  \BibitemOpen
  \bibfield  {author} {\bibinfo {author} {\bibfnamefont {G.}~\bibnamefont
  {Aad}} \emph {et~al.} (\bibinfo {collaboration} {ATLAS Collaboration}),\
  }\href@noop {} {\  (\bibinfo {year} {2014}{\natexlab{a}})},\ \bibinfo {note}
  {{A}TLAS-CONF-2014-010}\BibitemShut {NoStop}%
%%CITATION = ATLAS-CONF-2014-010 ETC.;%%
\bibitem [{\citenamefont {Aad}\ \emph {et~al.}(2013{\natexlab{g}})\citenamefont
  {Aad} \emph {et~al.}}]{ATLAS:2013mla}%
  \BibitemOpen
  \bibfield  {author} {\bibinfo {author} {\bibfnamefont {G.}~\bibnamefont
  {Aad}} \emph {et~al.} (\bibinfo {collaboration} {ATLAS Collaboration}),\
  }\href@noop {} {\  (\bibinfo {year} {2013}{\natexlab{g}})},\ \bibinfo {note}
  {{A}TLAS-CONF-2013-040}\BibitemShut {NoStop}%
%%CITATION = ATLAS-CONF-2013-040 ETC.;%%
\bibitem [{\citenamefont {Aad}\ \emph {et~al.}(2013{\natexlab{h}})\citenamefont
  {Aad} \emph {et~al.}}]{Aad:2013xqa}%
  \BibitemOpen
  \bibfield  {author} {\bibinfo {author} {\bibfnamefont {G.}~\bibnamefont
  {Aad}} \emph {et~al.} (\bibinfo {collaboration} {ATLAS Collaboration}),\
  }\href {\doibase 10.1016/j.physletb.2013.08.026} {\bibfield  {journal}
  {\bibinfo  {journal} {Phys.Lett.}\ }\textbf {\bibinfo {volume} {B726}},\
  \bibinfo {pages} {120} (\bibinfo {year} {2013}{\natexlab{h}})},\ \Eprint
  {http://arxiv.org/abs/1307.1432} {arXiv:1307.1432 [hep-ex]} \BibitemShut
  {NoStop}%
%%CITATION = ARXIV:1307.1432;%%
\bibitem [{\citenamefont {Aad}\ \emph {et~al.}(2013{\natexlab{i}})\citenamefont
  {Aad} \emph {et~al.}}]{ATLAS:2013xla}%
  \BibitemOpen
  \bibfield  {author} {\bibinfo {author} {\bibfnamefont {G.}~\bibnamefont
  {Aad}} \emph {et~al.} (\bibinfo {collaboration} {ATLAS Collaboration}),\
  }\href@noop {} {\  (\bibinfo {year} {2013}{\natexlab{i}})},\ \bibinfo {note}
  {{A}TLAS-CONF-2013-029}\BibitemShut {NoStop}%
%%CITATION = ATLAS-CONF-2013-029 ETC.;%%
\bibitem [{\citenamefont {Aad}\ \emph {et~al.}(2013{\natexlab{j}})\citenamefont
  {Aad} \emph {et~al.}}]{ATLAS:2013vla}%
  \BibitemOpen
  \bibfield  {author} {\bibinfo {author} {\bibfnamefont {G.}~\bibnamefont
  {Aad}} \emph {et~al.} (\bibinfo {collaboration} {ATLAS Collaboration}),\
  }\href@noop {} {\  (\bibinfo {year} {2013}{\natexlab{j}})},\ \bibinfo {note}
  {{A}TLAS-CONF-2013-031}\BibitemShut {NoStop}%
%%CITATION = ATLAS-CONF-2013-031 ETC.;%%
\bibitem [{\citenamefont {Chatrchyan}\ \emph
  {et~al.}(2013{\natexlab{e}})\citenamefont {Chatrchyan} \emph
  {et~al.}}]{Chatrchyan:2013iaa}%
  \BibitemOpen
  \bibfield  {author} {\bibinfo {author} {\bibfnamefont {S.}~\bibnamefont
  {Chatrchyan}} \emph {et~al.} (\bibinfo {collaboration} {CMS Collaboration}),\
  }\href@noop {} {\  (\bibinfo {year} {2013}{\natexlab{e}})},\ \Eprint
  {http://arxiv.org/abs/1312.1129} {arXiv:1312.1129 [hep-ex]} \BibitemShut
  {NoStop}%
%%CITATION = ARXIV:1312.1129;%%
\bibitem [{\citenamefont {Chatrchyan}\ \emph
  {et~al.}(2013{\natexlab{f}})\citenamefont {Chatrchyan} \emph
  {et~al.}}]{CMS:2013wda}%
  \BibitemOpen
  \bibfield  {author} {\bibinfo {author} {\bibfnamefont {S.}~\bibnamefont
  {Chatrchyan}} \emph {et~al.} (\bibinfo {collaboration} {CMS Collaboration}),\
  }\href@noop {} {\  (\bibinfo {year} {2013}{\natexlab{f}})},\ \bibinfo {note}
  {{C}MS-PAS-HIG-13-016}\BibitemShut {NoStop}%
%%CITATION = CMS-PAS-HIG-13-016 ETC.;%%
\bibitem [{\citenamefont {Aad}\ \emph {et~al.}(2013{\natexlab{k}})\citenamefont
  {Aad} \emph {et~al.}}]{ATLAS:tautau}%
  \BibitemOpen
  \bibfield  {author} {\bibinfo {author} {\bibfnamefont {G.}~\bibnamefont
  {Aad}} \emph {et~al.} (\bibinfo {collaboration} {ATLAS Collaboration}),\
  }\href@noop {} {\  (\bibinfo {year} {2013}{\natexlab{k}})},\ \bibinfo {note}
  {{A}TLAS-CONF-2013-108}\BibitemShut {NoStop}%
%%CITATION = ATLAS-CONF-2013-108 ETC.;%%
\bibitem [{\citenamefont {Chatrchyan}\ \emph
  {et~al.}(2014{\natexlab{a}})\citenamefont {Chatrchyan} \emph
  {et~al.}}]{Chatrchyan:2014nva}%
  \BibitemOpen
  \bibfield  {author} {\bibinfo {author} {\bibfnamefont {S.}~\bibnamefont
  {Chatrchyan}} \emph {et~al.} (\bibinfo {collaboration} {CMS Collaboration}),\
  }\href@noop {} {\  (\bibinfo {year} {2014}{\natexlab{a}})},\ \Eprint
  {http://arxiv.org/abs/1401.5041} {arXiv:1401.5041 [hep-ex]} \BibitemShut
  {NoStop}%
%%CITATION = ARXIV:1401.5041;%%
\bibitem [{\citenamefont {Li}\ \emph {et~al.}(2013)\citenamefont {Li},
  \citenamefont {Si}, \citenamefont {Yang}, \citenamefont {Yang},\ and\
  \citenamefont {Zheng}}]{Li:2013ava}%
  \BibitemOpen
  \bibfield  {author} {\bibinfo {author} {\bibfnamefont {H.-L.}\ \bibnamefont
  {Li}}, \bibinfo {author} {\bibfnamefont {Z.-G.}\ \bibnamefont {Si}}, \bibinfo
  {author} {\bibfnamefont {X.-Y.}\ \bibnamefont {Yang}}, \bibinfo {author}
  {\bibfnamefont {Z.-J.}\ \bibnamefont {Yang}}, \ and\ \bibinfo {author}
  {\bibfnamefont {Y.-J.}\ \bibnamefont {Zheng}},\ }\href {\doibase
  10.1103/PhysRevD.87.115024} {\bibfield  {journal} {\bibinfo  {journal}
  {Phys.Rev.}\ }\textbf {\bibinfo {volume} {D87}},\ \bibinfo {pages} {115024}
  (\bibinfo {year} {2013})},\ \Eprint {http://arxiv.org/abs/1305.1457}
  {arXiv:1305.1457 [hep-ph]} \BibitemShut {NoStop}%
%%CITATION = ARXIV:1305.1457;%%
\bibitem [{\citenamefont {Rentala}\ \emph {et~al.}(2013)\citenamefont
  {Rentala}, \citenamefont {Vignaroli}, \citenamefont {Li}, \citenamefont
  {Li},\ and\ \citenamefont {Yuan}}]{Rentala:2013uaa}%
  \BibitemOpen
  \bibfield  {author} {\bibinfo {author} {\bibfnamefont {V.}~\bibnamefont
  {Rentala}}, \bibinfo {author} {\bibfnamefont {N.}~\bibnamefont {Vignaroli}},
  \bibinfo {author} {\bibfnamefont {H.-n.}\ \bibnamefont {Li}}, \bibinfo
  {author} {\bibfnamefont {Z.}~\bibnamefont {Li}}, \ and\ \bibinfo {author}
  {\bibfnamefont {C.~P.}\ \bibnamefont {Yuan}},\ }\href {\doibase
  10.1103/PhysRevD.88.073007} {\bibfield  {journal} {\bibinfo  {journal}
  {Phys.Rev.}\ }\textbf {\bibinfo {volume} {D88}},\ \bibinfo {pages} {073007}
  (\bibinfo {year} {2013})},\ \Eprint {http://arxiv.org/abs/1306.0899}
  {arXiv:1306.0899 [hep-ph]} \BibitemShut {NoStop}%
%%CITATION = ARXIV:1306.0899;%%
\bibitem [{\citenamefont {Isidori}\ and\ \citenamefont
  {Trott}(2013)}]{Isidori:2013cga}%
  \BibitemOpen
  \bibfield  {author} {\bibinfo {author} {\bibfnamefont {G.}~\bibnamefont
  {Isidori}}\ and\ \bibinfo {author} {\bibfnamefont {M.}~\bibnamefont
  {Trott}},\ }\href@noop {} {\  (\bibinfo {year} {2013})},\ \Eprint
  {http://arxiv.org/abs/1307.4051} {arXiv:1307.4051 [hep-ph]} \BibitemShut
  {NoStop}%
%%CITATION = ARXIV:1307.4051;%%
\bibitem [{\citenamefont {Banerjee}\ \emph {et~al.}(2013)\citenamefont
  {Banerjee}, \citenamefont {Mukhopadhyay},\ and\ \citenamefont
  {Mukhopadhyaya}}]{Banerjee:2013apa}%
  \BibitemOpen
  \bibfield  {author} {\bibinfo {author} {\bibfnamefont {S.}~\bibnamefont
  {Banerjee}}, \bibinfo {author} {\bibfnamefont {S.}~\bibnamefont
  {Mukhopadhyay}}, \ and\ \bibinfo {author} {\bibfnamefont {B.}~\bibnamefont
  {Mukhopadhyaya}},\ }\href@noop {} {\  (\bibinfo {year} {2013})},\ \Eprint
  {http://arxiv.org/abs/1308.4860} {arXiv:1308.4860 [hep-ph]} \BibitemShut
  {NoStop}%
%%CITATION = ARXIV:1308.4860;%%
\bibitem [{\citenamefont {Gao}(2013)}]{Gao:2013nga}%
  \BibitemOpen
  \bibfield  {author} {\bibinfo {author} {\bibfnamefont {J.}~\bibnamefont
  {Gao}},\ }\href@noop {} {\  (\bibinfo {year} {2013})},\ \Eprint
  {http://arxiv.org/abs/1308.5453} {arXiv:1308.5453 [hep-ph]} \BibitemShut
  {NoStop}%
%%CITATION = ARXIV:1308.5453;%%
\bibitem [{\citenamefont {Azatov}\ and\ \citenamefont
  {Paul}(2013)}]{Azatov:2013xha}%
  \BibitemOpen
  \bibfield  {author} {\bibinfo {author} {\bibfnamefont {A.}~\bibnamefont
  {Azatov}}\ and\ \bibinfo {author} {\bibfnamefont {A.}~\bibnamefont {Paul}},\
  }\href@noop {} {\  (\bibinfo {year} {2013})},\ \Eprint
  {http://arxiv.org/abs/1309.5273} {arXiv:1309.5273 [hep-ph]} \BibitemShut
  {NoStop}%
%%CITATION = ARXIV:1309.5273;%%
\bibitem [{\citenamefont {Englert}\ \emph {et~al.}(2013)\citenamefont
  {Englert}, \citenamefont {McCullough},\ and\ \citenamefont
  {Spannowsky}}]{Englert:2013vua}%
  \BibitemOpen
  \bibfield  {author} {\bibinfo {author} {\bibfnamefont {C.}~\bibnamefont
  {Englert}}, \bibinfo {author} {\bibfnamefont {M.}~\bibnamefont {McCullough}},
  \ and\ \bibinfo {author} {\bibfnamefont {M.}~\bibnamefont {Spannowsky}},\
  }\href@noop {} {\  (\bibinfo {year} {2013})},\ \Eprint
  {http://arxiv.org/abs/1310.4828} {arXiv:1310.4828 [hep-ph]} \BibitemShut
  {NoStop}%
%%CITATION = ARXIV:1310.4828;%%
\bibitem [{\citenamefont {Maltoni}\ \emph {et~al.}(2013)\citenamefont
  {Maltoni}, \citenamefont {Mawatari},\ and\ \citenamefont
  {Zaro}}]{Maltoni:2013sma}%
  \BibitemOpen
  \bibfield  {author} {\bibinfo {author} {\bibfnamefont {F.}~\bibnamefont
  {Maltoni}}, \bibinfo {author} {\bibfnamefont {K.}~\bibnamefont {Mawatari}}, \
  and\ \bibinfo {author} {\bibfnamefont {M.}~\bibnamefont {Zaro}},\ }\href@noop
  {} {\  (\bibinfo {year} {2013})},\ \Eprint {http://arxiv.org/abs/1311.1829}
  {arXiv:1311.1829 [hep-ph]} \BibitemShut {NoStop}%
%%CITATION = ARXIV:1311.1829;%%
\bibitem [{\citenamefont {Huang}\ \emph {et~al.}(2013)\citenamefont {Huang},
  \citenamefont {Liu}, \citenamefont {Wang},\ and\ \citenamefont
  {Yu}}]{Huang:2013ima}%
  \BibitemOpen
  \bibfield  {author} {\bibinfo {author} {\bibfnamefont {J.}~\bibnamefont
  {Huang}}, \bibinfo {author} {\bibfnamefont {T.}~\bibnamefont {Liu}}, \bibinfo
  {author} {\bibfnamefont {L.-T.}\ \bibnamefont {Wang}}, \ and\ \bibinfo
  {author} {\bibfnamefont {F.}~\bibnamefont {Yu}},\ }\href@noop {} {\
  (\bibinfo {year} {2013})},\ \Eprint {http://arxiv.org/abs/1309.6633}
  {arXiv:1309.6633 [hep-ph]} \BibitemShut {NoStop}%
%%CITATION = ARXIV:1309.6633;%%
\bibitem [{\citenamefont {Curtin}\ \emph {et~al.}(2013)\citenamefont {Curtin},
  \citenamefont {Essig}, \citenamefont {Gori}, \citenamefont {Jaiswal},
  \citenamefont {Katz} \emph {et~al.}}]{Curtin:2013fra}%
  \BibitemOpen
  \bibfield  {author} {\bibinfo {author} {\bibfnamefont {D.}~\bibnamefont
  {Curtin}}, \bibinfo {author} {\bibfnamefont {R.}~\bibnamefont {Essig}},
  \bibinfo {author} {\bibfnamefont {S.}~\bibnamefont {Gori}}, \bibinfo {author}
  {\bibfnamefont {P.}~\bibnamefont {Jaiswal}}, \bibinfo {author} {\bibfnamefont
  {A.}~\bibnamefont {Katz}},  \emph {et~al.},\ }\href@noop {} {\  (\bibinfo
  {year} {2013})},\ \Eprint {http://arxiv.org/abs/1312.4992} {arXiv:1312.4992
  [hep-ph]} \BibitemShut {NoStop}%
%%CITATION = ARXIV:1312.4992;%%
\bibitem [{\citenamefont {Aad}\ \emph {et~al.}(2013{\natexlab{l}})\citenamefont
  {Aad} \emph {et~al.}}]{ATLAS:2013pma}%
  \BibitemOpen
  \bibfield  {author} {\bibinfo {author} {\bibfnamefont {G.}~\bibnamefont
  {Aad}} \emph {et~al.} (\bibinfo {collaboration} {ATLAS Collaboration}),\
  }\href@noop {} {\  (\bibinfo {year} {2013}{\natexlab{l}})},\ \bibinfo {note}
  {{A}TLAS-CONF-2013-011}\BibitemShut {NoStop}%
%%CITATION = ATLAS-CONF-2013-011 ETC.;%%
\bibitem [{\citenamefont {Chatrchyan}\ \emph
  {et~al.}(2013{\natexlab{g}})\citenamefont {Chatrchyan} \emph
  {et~al.}}]{CMS:1900fga}%
  \BibitemOpen
  \bibfield  {author} {\bibinfo {author} {\bibfnamefont {S.}~\bibnamefont
  {Chatrchyan}} \emph {et~al.} (\bibinfo {collaboration} {CMS Collaboration}),\
  }\href@noop {} {\  (\bibinfo {year} {2013}{\natexlab{g}})},\ \bibinfo {note}
  {{C}MS-PAS-HIG-13-028}\BibitemShut {NoStop}%
%%CITATION = CMS-PAS-HIG-13-028 ETC.;%%
\bibitem [{\citenamefont {Chatrchyan}\ \emph
  {et~al.}(2013{\natexlab{h}})\citenamefont {Chatrchyan} \emph
  {et~al.}}]{CMS:2013bfa}%
  \BibitemOpen
  \bibfield  {author} {\bibinfo {author} {\bibfnamefont {S.}~\bibnamefont
  {Chatrchyan}} \emph {et~al.} (\bibinfo {collaboration} {CMS Collaboration}),\
  }\href@noop {} {\  (\bibinfo {year} {2013}{\natexlab{h}})},\ \bibinfo {note}
  {{C}MS-PAS-HIG-13-013}\BibitemShut {NoStop}%
%%CITATION = CMS-PAS-HIG-13-013 ETC.;%%
\bibitem [{\citenamefont {Chatrchyan}\ \emph
  {et~al.}(2013{\natexlab{i}})\citenamefont {Chatrchyan} \emph
  {et~al.}}]{CMS:2013yda}%
  \BibitemOpen
  \bibfield  {author} {\bibinfo {author} {\bibfnamefont {S.}~\bibnamefont
  {Chatrchyan}} \emph {et~al.} (\bibinfo {collaboration} {CMS Collaboration}),\
  }\href@noop {} {\  (\bibinfo {year} {2013}{\natexlab{i}})},\ \bibinfo {note}
  {{C}MS-PAS-HIG-13-018}\BibitemShut {NoStop}%
%%CITATION = CMS-PAS-HIG-13-018 ETC.;%%
\bibitem [{\citenamefont {Caola}\ and\ \citenamefont
  {Melnikov}(2013)}]{Caola:2013yja}%
  \BibitemOpen
  \bibfield  {author} {\bibinfo {author} {\bibfnamefont {F.}~\bibnamefont
  {Caola}}\ and\ \bibinfo {author} {\bibfnamefont {K.}~\bibnamefont
  {Melnikov}},\ }\href {\doibase 10.1103/PhysRevD.88.054024} {\bibfield
  {journal} {\bibinfo  {journal} {Phys.Rev.}\ }\textbf {\bibinfo {volume}
  {D88}},\ \bibinfo {pages} {054024} (\bibinfo {year} {2013})},\ \Eprint
  {http://arxiv.org/abs/1307.4935} {arXiv:1307.4935 [hep-ph]} \BibitemShut
  {NoStop}%
%%CITATION = ARXIV:1307.4935;%%
\bibitem [{\citenamefont {Campbell}\ \emph {et~al.}(2013)\citenamefont
  {Campbell}, \citenamefont {Ellis},\ and\ \citenamefont
  {Williams}}]{Campbell:2013una}%
  \BibitemOpen
  \bibfield  {author} {\bibinfo {author} {\bibfnamefont {J.~M.}\ \bibnamefont
  {Campbell}}, \bibinfo {author} {\bibfnamefont {R.~K.}\ \bibnamefont {Ellis}},
  \ and\ \bibinfo {author} {\bibfnamefont {C.}~\bibnamefont {Williams}},\
  }\href@noop {} {\  (\bibinfo {year} {2013})},\ \Eprint
  {http://arxiv.org/abs/1311.3589} {arXiv:1311.3589 [hep-ph]} \BibitemShut
  {NoStop}%
%%CITATION = ARXIV:1311.3589;%%
\bibitem [{\citenamefont {Campbell}\ \emph {et~al.}(2014)\citenamefont
  {Campbell}, \citenamefont {Ellis},\ and\ \citenamefont
  {Williams}}]{Campbell:2013wga}%
  \BibitemOpen
  \bibfield  {author} {\bibinfo {author} {\bibfnamefont {J.~M.}\ \bibnamefont
  {Campbell}}, \bibinfo {author} {\bibfnamefont {R.~K.}\ \bibnamefont {Ellis}},
  \ and\ \bibinfo {author} {\bibfnamefont {C.}~\bibnamefont {Williams}},\
  }\href {\doibase 10.1103/PhysRevD.89.053011} {\bibfield  {journal} {\bibinfo
  {journal} {Phys.Rev.}\ }\textbf {\bibinfo {volume} {D89}},\ \bibinfo {pages}
  {053011} (\bibinfo {year} {2014})},\ \Eprint {http://arxiv.org/abs/1312.1628}
  {arXiv:1312.1628 [hep-ph]} \BibitemShut {NoStop}%
%%CITATION = ARXIV:1312.1628;%%
\bibitem [{\citenamefont {Chatrchyan}\ \emph
  {et~al.}(2014{\natexlab{b}})\citenamefont {Chatrchyan} \emph
  {et~al.}}]{CMS:2014ala}%
  \BibitemOpen
  \bibfield  {author} {\bibinfo {author} {\bibfnamefont {S.}~\bibnamefont
  {Chatrchyan}} \emph {et~al.} (\bibinfo {collaboration} {CMS Collaboration}),\
  }\href@noop {} {\  (\bibinfo {year} {2014}{\natexlab{b}})},\ \bibinfo {note}
  {{C}MS-PAS-HIG-14-002}\BibitemShut {NoStop}%
%%CITATION = CMS-PAS-HIG-14-002 ETC.;%%
\bibitem [{\citenamefont {Chatrchyan}\ \emph
  {et~al.}(2013{\natexlab{j}})\citenamefont {Chatrchyan} \emph
  {et~al.}}]{CMS:2013eua}%
  \BibitemOpen
  \bibfield  {author} {\bibinfo {author} {\bibfnamefont {S.}~\bibnamefont
  {Chatrchyan}} \emph {et~al.} (\bibinfo {collaboration} {CMS Collaboration}),\
  }\href@noop {} {\  (\bibinfo {year} {2013}{\natexlab{j}})},\ \bibinfo {note}
  {{C}MS-PAS-HIG-13-025}\BibitemShut {NoStop}%
%%CITATION = CMS-PAS-HIG-13-025 ETC.;%%
\bibitem [{\citenamefont {Datta}\ \emph {et~al.}(2004)\citenamefont {Datta},
  \citenamefont {Djouadi}, \citenamefont {Guchait},\ and\ \citenamefont
  {Moortgat}}]{Datta:2003iz}%
  \BibitemOpen
  \bibfield  {author} {\bibinfo {author} {\bibfnamefont {A.}~\bibnamefont
  {Datta}}, \bibinfo {author} {\bibfnamefont {A.}~\bibnamefont {Djouadi}},
  \bibinfo {author} {\bibfnamefont {M.}~\bibnamefont {Guchait}}, \ and\
  \bibinfo {author} {\bibfnamefont {F.}~\bibnamefont {Moortgat}},\ }\href
  {\doibase 10.1016/j.nuclphysb.2003.12.012} {\bibfield  {journal} {\bibinfo
  {journal} {Nucl.Phys.}\ }\textbf {\bibinfo {volume} {B681}},\ \bibinfo
  {pages} {31} (\bibinfo {year} {2004})},\ \Eprint
  {http://arxiv.org/abs/hep-ph/0303095} {arXiv:hep-ph/0303095 [hep-ph]}
  \BibitemShut {NoStop}%
%%CITATION = HEP-PH/0303095;%%
\bibitem [{\citenamefont {Huitu}\ \emph {et~al.}(2008)\citenamefont {Huitu},
  \citenamefont {Kinnunen}, \citenamefont {Laamanen}, \citenamefont {Lehti},
  \citenamefont {Roy} \emph {et~al.}}]{Huitu:2008sa}%
  \BibitemOpen
  \bibfield  {author} {\bibinfo {author} {\bibfnamefont {K.}~\bibnamefont
  {Huitu}}, \bibinfo {author} {\bibfnamefont {R.}~\bibnamefont {Kinnunen}},
  \bibinfo {author} {\bibfnamefont {J.}~\bibnamefont {Laamanen}}, \bibinfo
  {author} {\bibfnamefont {S.}~\bibnamefont {Lehti}}, \bibinfo {author}
  {\bibfnamefont {S.}~\bibnamefont {Roy}},  \emph {et~al.},\ }\href {\doibase
  10.1140/epjc/s10052-008-0786-0} {\bibfield  {journal} {\bibinfo  {journal}
  {Eur.Phys.J.}\ }\textbf {\bibinfo {volume} {C58}},\ \bibinfo {pages} {591}
  (\bibinfo {year} {2008})},\ \Eprint {http://arxiv.org/abs/0808.3094}
  {arXiv:0808.3094 [hep-ph]} \BibitemShut {NoStop}%
%%CITATION = ARXIV:0808.3094;%%
\bibitem [{\citenamefont {Gori}\ \emph {et~al.}(2011)\citenamefont {Gori},
  \citenamefont {Schwaller},\ and\ \citenamefont {Wagner}}]{Gori:2011hj}%
  \BibitemOpen
  \bibfield  {author} {\bibinfo {author} {\bibfnamefont {S.}~\bibnamefont
  {Gori}}, \bibinfo {author} {\bibfnamefont {P.}~\bibnamefont {Schwaller}}, \
  and\ \bibinfo {author} {\bibfnamefont {C.~E.}\ \bibnamefont {Wagner}},\
  }\href {\doibase 10.1103/PhysRevD.83.115022} {\bibfield  {journal} {\bibinfo
  {journal} {Phys.Rev.}\ }\textbf {\bibinfo {volume} {D83}},\ \bibinfo {pages}
  {115022} (\bibinfo {year} {2011})},\ \Eprint {http://arxiv.org/abs/1103.4138}
  {arXiv:1103.4138 [hep-ph]} \BibitemShut {NoStop}%
%%CITATION = ARXIV:1103.4138;%%
\bibitem [{\citenamefont {Stal}\ and\ \citenamefont
  {Weiglein}(2012)}]{Stal:2011cz}%
  \BibitemOpen
  \bibfield  {author} {\bibinfo {author} {\bibfnamefont {O.}~\bibnamefont
  {Stal}}\ and\ \bibinfo {author} {\bibfnamefont {G.}~\bibnamefont
  {Weiglein}},\ }\href {\doibase 10.1007/JHEP01(2012)071} {\bibfield  {journal}
  {\bibinfo  {journal} {JHEP}\ }\textbf {\bibinfo {volume} {1201}},\ \bibinfo
  {pages} {071} (\bibinfo {year} {2012})},\ \Eprint
  {http://arxiv.org/abs/1108.0595} {arXiv:1108.0595 [hep-ph]} \BibitemShut
  {NoStop}%
%%CITATION = ARXIV:1108.0595;%%
\bibitem [{\citenamefont {Butterworth}\ \emph {et~al.}(2008)\citenamefont
  {Butterworth}, \citenamefont {Davison}, \citenamefont {Rubin},\ and\
  \citenamefont {Salam}}]{Butterworth:2008iy}%
  \BibitemOpen
  \bibfield  {author} {\bibinfo {author} {\bibfnamefont {J.~M.}\ \bibnamefont
  {Butterworth}}, \bibinfo {author} {\bibfnamefont {A.~R.}\ \bibnamefont
  {Davison}}, \bibinfo {author} {\bibfnamefont {M.}~\bibnamefont {Rubin}}, \
  and\ \bibinfo {author} {\bibfnamefont {G.~P.}\ \bibnamefont {Salam}},\ }\href
  {\doibase 10.1103/PhysRevLett.100.242001} {\bibfield  {journal} {\bibinfo
  {journal} {Phys.Rev.Lett.}\ }\textbf {\bibinfo {volume} {100}},\ \bibinfo
  {pages} {242001} (\bibinfo {year} {2008})},\ \Eprint
  {http://arxiv.org/abs/0802.2470} {arXiv:0802.2470 [hep-ph]} \BibitemShut
  {NoStop}%
%%CITATION = ARXIV:0802.2470;%%
\bibitem [{\citenamefont {Kribs}\ \emph
  {et~al.}(2010{\natexlab{a}})\citenamefont {Kribs}, \citenamefont {Martin},
  \citenamefont {Roy},\ and\ \citenamefont {Spannowsky}}]{Kribs:2009yh}%
  \BibitemOpen
  \bibfield  {author} {\bibinfo {author} {\bibfnamefont {G.~D.}\ \bibnamefont
  {Kribs}}, \bibinfo {author} {\bibfnamefont {A.}~\bibnamefont {Martin}},
  \bibinfo {author} {\bibfnamefont {T.~S.}\ \bibnamefont {Roy}}, \ and\
  \bibinfo {author} {\bibfnamefont {M.}~\bibnamefont {Spannowsky}},\ }\href
  {\doibase 10.1103/PhysRevD.81.111501} {\bibfield  {journal} {\bibinfo
  {journal} {Phys.Rev.}\ }\textbf {\bibinfo {volume} {D81}},\ \bibinfo {pages}
  {111501} (\bibinfo {year} {2010}{\natexlab{a}})},\ \Eprint
  {http://arxiv.org/abs/0912.4731} {arXiv:0912.4731 [hep-ph]} \BibitemShut
  {NoStop}%
%%CITATION = ARXIV:0912.4731;%%
\bibitem [{\citenamefont {Kribs}\ \emph
  {et~al.}(2010{\natexlab{b}})\citenamefont {Kribs}, \citenamefont {Martin},
  \citenamefont {Roy},\ and\ \citenamefont {Spannowsky}}]{Kribs:2010hp}%
  \BibitemOpen
  \bibfield  {author} {\bibinfo {author} {\bibfnamefont {G.~D.}\ \bibnamefont
  {Kribs}}, \bibinfo {author} {\bibfnamefont {A.}~\bibnamefont {Martin}},
  \bibinfo {author} {\bibfnamefont {T.~S.}\ \bibnamefont {Roy}}, \ and\
  \bibinfo {author} {\bibfnamefont {M.}~\bibnamefont {Spannowsky}},\ }\href
  {\doibase 10.1103/PhysRevD.82.095012} {\bibfield  {journal} {\bibinfo
  {journal} {Phys.Rev.}\ }\textbf {\bibinfo {volume} {D82}},\ \bibinfo {pages}
  {095012} (\bibinfo {year} {2010}{\natexlab{b}})},\ \Eprint
  {http://arxiv.org/abs/1006.1656} {arXiv:1006.1656 [hep-ph]} \BibitemShut
  {NoStop}%
%%CITATION = ARXIV:1006.1656;%%
\bibitem [{\citenamefont {Gunion}\ \emph {et~al.}(1987)\citenamefont {Gunion},
  \citenamefont {Haber}, \citenamefont {Barnett}, \citenamefont {Drees},
  \citenamefont {Karatas} \emph {et~al.}}]{Gunion:1987kg}%
  \BibitemOpen
  \bibfield  {author} {\bibinfo {author} {\bibfnamefont {J.}~\bibnamefont
  {Gunion}}, \bibinfo {author} {\bibfnamefont {H.}~\bibnamefont {Haber}},
  \bibinfo {author} {\bibfnamefont {R.~M.}\ \bibnamefont {Barnett}}, \bibinfo
  {author} {\bibfnamefont {M.}~\bibnamefont {Drees}}, \bibinfo {author}
  {\bibfnamefont {D.}~\bibnamefont {Karatas}},  \emph {et~al.},\ }\href
  {\doibase 10.1142/S0217751X87000533} {\bibfield  {journal} {\bibinfo
  {journal} {Int.J.Mod.Phys.}\ }\textbf {\bibinfo {volume} {A2}},\ \bibinfo
  {pages} {1145} (\bibinfo {year} {1987})}\BibitemShut {NoStop}%
%%CITATION = IMPAE,A2,1145;%%
\bibitem [{\citenamefont {Gunion}\ and\ \citenamefont
  {Haber}(1988)}]{Gunion:1987yh}%
  \BibitemOpen
  \bibfield  {author} {\bibinfo {author} {\bibfnamefont {J.~F.}\ \bibnamefont
  {Gunion}}\ and\ \bibinfo {author} {\bibfnamefont {H.~E.}\ \bibnamefont
  {Haber}},\ }\href {\doibase 10.1103/PhysRevD.37.2515} {\bibfield  {journal}
  {\bibinfo  {journal} {Phys.Rev.}\ }\textbf {\bibinfo {volume} {D37}},\
  \bibinfo {pages} {2515} (\bibinfo {year} {1988})}\BibitemShut {NoStop}%
%%CITATION = PHRVA,D37,2515;%%
\bibitem [{\citenamefont {Bartl}\ \emph {et~al.}(1989)\citenamefont {Bartl},
  \citenamefont {Majerotto},\ and\ \citenamefont {Oshimo}}]{Bartl:1988cn}%
  \BibitemOpen
  \bibfield  {author} {\bibinfo {author} {\bibfnamefont {A.}~\bibnamefont
  {Bartl}}, \bibinfo {author} {\bibfnamefont {W.}~\bibnamefont {Majerotto}}, \
  and\ \bibinfo {author} {\bibfnamefont {N.}~\bibnamefont {Oshimo}},\ }\href
  {\doibase 10.1016/0370-2693(89)91401-9} {\bibfield  {journal} {\bibinfo
  {journal} {Phys.Lett.}\ }\textbf {\bibinfo {volume} {B216}},\ \bibinfo
  {pages} {233} (\bibinfo {year} {1989})}\BibitemShut {NoStop}%
%%CITATION = PHLTA,B216,233;%%
\bibitem [{\citenamefont {Djouadi}\ \emph {et~al.}(2001)\citenamefont
  {Djouadi}, \citenamefont {Mambrini},\ and\ \citenamefont
  {Muhlleitner}}]{Djouadi:2001fa}%
  \BibitemOpen
  \bibfield  {author} {\bibinfo {author} {\bibfnamefont {A.}~\bibnamefont
  {Djouadi}}, \bibinfo {author} {\bibfnamefont {Y.}~\bibnamefont {Mambrini}}, \
  and\ \bibinfo {author} {\bibfnamefont {M.}~\bibnamefont {Muhlleitner}},\
  }\href {\doibase 10.1007/s100520100679} {\bibfield  {journal} {\bibinfo
  {journal} {Eur.Phys.J.}\ }\textbf {\bibinfo {volume} {C20}},\ \bibinfo
  {pages} {563} (\bibinfo {year} {2001})},\ \Eprint
  {http://arxiv.org/abs/hep-ph/0104115} {arXiv:hep-ph/0104115 [hep-ph]}
  \BibitemShut {NoStop}%
%%CITATION = HEP-PH/0104115;%%
\bibitem [{\citenamefont {Diaz}\ and\ \citenamefont
  {Fileviez~Perez}(2005)}]{Diaz:2004qt}%
  \BibitemOpen
  \bibfield  {author} {\bibinfo {author} {\bibfnamefont {M.~A.}\ \bibnamefont
  {Diaz}}\ and\ \bibinfo {author} {\bibfnamefont {P.}~\bibnamefont
  {Fileviez~Perez}},\ }\href {\doibase 10.1088/0954-3899/31/7/003} {\bibfield
  {journal} {\bibinfo  {journal} {J.Phys.}\ }\textbf {\bibinfo {volume}
  {G31}},\ \bibinfo {pages} {563} (\bibinfo {year} {2005})},\ \Eprint
  {http://arxiv.org/abs/hep-ph/0412066} {arXiv:hep-ph/0412066 [hep-ph]}
  \BibitemShut {NoStop}%
%%CITATION = HEP-PH/0412066;%%
\bibitem [{\citenamefont {Baer}\ \emph {et~al.}(2012)\citenamefont {Baer},
  \citenamefont {Barger}, \citenamefont {Lessa}, \citenamefont {Sreethawong},\
  and\ \citenamefont {Tata}}]{Baer:2012ts}%
  \BibitemOpen
  \bibfield  {author} {\bibinfo {author} {\bibfnamefont {H.}~\bibnamefont
  {Baer}}, \bibinfo {author} {\bibfnamefont {V.}~\bibnamefont {Barger}},
  \bibinfo {author} {\bibfnamefont {A.}~\bibnamefont {Lessa}}, \bibinfo
  {author} {\bibfnamefont {W.}~\bibnamefont {Sreethawong}}, \ and\ \bibinfo
  {author} {\bibfnamefont {X.}~\bibnamefont {Tata}},\ }\href {\doibase
  10.1103/PhysRevD.85.055022} {\bibfield  {journal} {\bibinfo  {journal}
  {Phys.Rev.}\ }\textbf {\bibinfo {volume} {D85}},\ \bibinfo {pages} {055022}
  (\bibinfo {year} {2012})},\ \Eprint {http://arxiv.org/abs/1201.2949}
  {arXiv:1201.2949 [hep-ph]} \BibitemShut {NoStop}%
%%CITATION = ARXIV:1201.2949;%%
\bibitem [{\citenamefont {Ghosh}\ \emph {et~al.}(2012)\citenamefont {Ghosh},
  \citenamefont {Guchait},\ and\ \citenamefont {Sengupta}}]{Ghosh:2012mc}%
  \BibitemOpen
  \bibfield  {author} {\bibinfo {author} {\bibfnamefont {D.}~\bibnamefont
  {Ghosh}}, \bibinfo {author} {\bibfnamefont {M.}~\bibnamefont {Guchait}}, \
  and\ \bibinfo {author} {\bibfnamefont {D.}~\bibnamefont {Sengupta}},\ }\href
  {\doibase 10.1140/epjc/s10052-012-2141-8} {\bibfield  {journal} {\bibinfo
  {journal} {Eur.Phys.J.}\ }\textbf {\bibinfo {volume} {C72}},\ \bibinfo
  {pages} {2141} (\bibinfo {year} {2012})},\ \Eprint
  {http://arxiv.org/abs/1202.4937} {arXiv:1202.4937 [hep-ph]} \BibitemShut
  {NoStop}%
%%CITATION = ARXIV:1202.4937;%%
\bibitem [{\citenamefont {Arbey}\ \emph {et~al.}(2012)\citenamefont {Arbey},
  \citenamefont {Battaglia},\ and\ \citenamefont {Mahmoudi}}]{Arbey:2012fa}%
  \BibitemOpen
  \bibfield  {author} {\bibinfo {author} {\bibfnamefont {A.}~\bibnamefont
  {Arbey}}, \bibinfo {author} {\bibfnamefont {M.}~\bibnamefont {Battaglia}}, \
  and\ \bibinfo {author} {\bibfnamefont {F.}~\bibnamefont {Mahmoudi}},\
  }\href@noop {} {\  (\bibinfo {year} {2012})},\ \Eprint
  {http://arxiv.org/abs/1212.6865} {arXiv:1212.6865 [hep-ph]} \BibitemShut
  {NoStop}%
%%CITATION = ARXIV:1212.6865;%%
\bibitem [{\citenamefont {Baer}\ \emph
  {et~al.}(2013{\natexlab{a}})\citenamefont {Baer}, \citenamefont {Barger},\
  and\ \citenamefont {Mickelson}}]{Baer:2013vpa}%
  \BibitemOpen
  \bibfield  {author} {\bibinfo {author} {\bibfnamefont {H.}~\bibnamefont
  {Baer}}, \bibinfo {author} {\bibfnamefont {V.}~\bibnamefont {Barger}}, \ and\
  \bibinfo {author} {\bibfnamefont {D.}~\bibnamefont {Mickelson}},\ }\href
  {\doibase 10.1016/j.physletb.2013.08.060} {\bibfield  {journal} {\bibinfo
  {journal} {Phys.Lett.}\ }\textbf {\bibinfo {volume} {B726}},\ \bibinfo
  {pages} {330} (\bibinfo {year} {2013}{\natexlab{a}})},\ \Eprint
  {http://arxiv.org/abs/1303.3816} {arXiv:1303.3816 [hep-ph]} \BibitemShut
  {NoStop}%
%%CITATION = ARXIV:1303.3816;%%
\bibitem [{\citenamefont {Baer}\ \emph
  {et~al.}(2013{\natexlab{b}})\citenamefont {Baer}, \citenamefont {Barger},
  \citenamefont {Huang}, \citenamefont {Mickelson}, \citenamefont {Mustafayev}
  \emph {et~al.}}]{Baer:2013ava}%
  \BibitemOpen
  \bibfield  {author} {\bibinfo {author} {\bibfnamefont {H.}~\bibnamefont
  {Baer}}, \bibinfo {author} {\bibfnamefont {V.}~\bibnamefont {Barger}},
  \bibinfo {author} {\bibfnamefont {P.}~\bibnamefont {Huang}}, \bibinfo
  {author} {\bibfnamefont {D.}~\bibnamefont {Mickelson}}, \bibinfo {author}
  {\bibfnamefont {A.}~\bibnamefont {Mustafayev}},  \emph {et~al.},\ }\href@noop
  {} {\  (\bibinfo {year} {2013}{\natexlab{b}})},\ \Eprint
  {http://arxiv.org/abs/1306.2926} {arXiv:1306.2926 [hep-ph]} \BibitemShut
  {NoStop}%
%%CITATION = ARXIV:1306.2926;%%
\bibitem [{\citenamefont {Baer}\ \emph
  {et~al.}(2013{\natexlab{c}})\citenamefont {Baer}, \citenamefont {Barger},
  \citenamefont {Huang}, \citenamefont {Mickelson}, \citenamefont {Mustafayev}
  \emph {et~al.}}]{Baer:2013faa}%
  \BibitemOpen
  \bibfield  {author} {\bibinfo {author} {\bibfnamefont {H.}~\bibnamefont
  {Baer}}, \bibinfo {author} {\bibfnamefont {V.}~\bibnamefont {Barger}},
  \bibinfo {author} {\bibfnamefont {P.}~\bibnamefont {Huang}}, \bibinfo
  {author} {\bibfnamefont {D.}~\bibnamefont {Mickelson}}, \bibinfo {author}
  {\bibfnamefont {A.}~\bibnamefont {Mustafayev}},  \emph {et~al.},\ }\href@noop
  {} {\  (\bibinfo {year} {2013}{\natexlab{c}})},\ \Eprint
  {http://arxiv.org/abs/1306.3148} {arXiv:1306.3148 [hep-ph]} \BibitemShut
  {NoStop}%
%%CITATION = ARXIV:1306.3148;%%
\bibitem [{\citenamefont {Han}\ \emph {et~al.}(2013{\natexlab{a}})\citenamefont
  {Han}, \citenamefont {Li}, \citenamefont {Su},\ and\ \citenamefont
  {Wang}}]{Han:2013mga}%
  \BibitemOpen
  \bibfield  {author} {\bibinfo {author} {\bibfnamefont {T.}~\bibnamefont
  {Han}}, \bibinfo {author} {\bibfnamefont {T.}~\bibnamefont {Li}}, \bibinfo
  {author} {\bibfnamefont {S.}~\bibnamefont {Su}}, \ and\ \bibinfo {author}
  {\bibfnamefont {L.-T.}\ \bibnamefont {Wang}},\ }\href {\doibase
  10.1007/JHEP11(2013)053} {\bibfield  {journal} {\bibinfo  {journal} {JHEP}\
  }\textbf {\bibinfo {volume} {1311}},\ \bibinfo {pages} {053} (\bibinfo {year}
  {2013}{\natexlab{a}})},\ \Eprint {http://arxiv.org/abs/1306.3229}
  {arXiv:1306.3229 [hep-ph]} \BibitemShut {NoStop}%
%%CITATION = ARXIV:1306.3229;%%
\bibitem [{\citenamefont {Baer}\ \emph
  {et~al.}(2013{\natexlab{d}})\citenamefont {Baer}, \citenamefont {Barger},
  \citenamefont {Mickelson},\ and\ \citenamefont {Tata}}]{Baer:2013ssa}%
  \BibitemOpen
  \bibfield  {author} {\bibinfo {author} {\bibfnamefont {H.}~\bibnamefont
  {Baer}}, \bibinfo {author} {\bibfnamefont {V.}~\bibnamefont {Barger}},
  \bibinfo {author} {\bibfnamefont {D.}~\bibnamefont {Mickelson}}, \ and\
  \bibinfo {author} {\bibfnamefont {X.}~\bibnamefont {Tata}},\ }\href@noop {}
  {\  (\bibinfo {year} {2013}{\natexlab{d}})},\ \Eprint
  {http://arxiv.org/abs/1306.4183} {arXiv:1306.4183 [hep-ph]} \BibitemShut
  {NoStop}%
%%CITATION = ARXIV:1306.4183;%%
\bibitem [{\citenamefont {Berggren}\ \emph
  {et~al.}(2013{\natexlab{a}})\citenamefont {Berggren}, \citenamefont
  {Brümmer}, \citenamefont {List}, \citenamefont {Moortgat-Pick},
  \citenamefont {Robens} \emph {et~al.}}]{Berggren:2013vfa}%
  \BibitemOpen
  \bibfield  {author} {\bibinfo {author} {\bibfnamefont {M.}~\bibnamefont
  {Berggren}}, \bibinfo {author} {\bibfnamefont {F.}~\bibnamefont {Brümmer}},
  \bibinfo {author} {\bibfnamefont {J.}~\bibnamefont {List}}, \bibinfo {author}
  {\bibfnamefont {G.}~\bibnamefont {Moortgat-Pick}}, \bibinfo {author}
  {\bibfnamefont {T.}~\bibnamefont {Robens}},  \emph {et~al.},\ }\href
  {\doibase 10.1140/epjc/s10052-013-2660-y} {\  (\bibinfo {year}
  {2013}{\natexlab{a}}),\ 10.1140/epjc/s10052-013-2660-y},\ \Eprint
  {http://arxiv.org/abs/1307.3566} {arXiv:1307.3566 [hep-ph]} \BibitemShut
  {NoStop}%
%%CITATION = ARXIV:1307.3566;%%
\bibitem [{\citenamefont {Bharucha}\ \emph {et~al.}(2013)\citenamefont
  {Bharucha}, \citenamefont {Heinemeyer},\ and\ \citenamefont
  {Pahlen}}]{Bharucha:2013epa}%
  \BibitemOpen
  \bibfield  {author} {\bibinfo {author} {\bibfnamefont {A.}~\bibnamefont
  {Bharucha}}, \bibinfo {author} {\bibfnamefont {S.}~\bibnamefont
  {Heinemeyer}}, \ and\ \bibinfo {author} {\bibfnamefont {F.}~\bibnamefont
  {Pahlen}},\ }\href {\doibase 10.1140/epjc/s10052-013-2629-x} {\bibfield
  {journal} {\bibinfo  {journal} {Eur.Phys.J.}\ }\textbf {\bibinfo {volume}
  {C73}},\ \bibinfo {pages} {2629} (\bibinfo {year} {2013})},\ \Eprint
  {http://arxiv.org/abs/1307.4237} {arXiv:1307.4237} \BibitemShut {NoStop}%
%%CITATION = ARXIV:1307.4237;%%
\bibitem [{\citenamefont {Gori}\ \emph {et~al.}(2013)\citenamefont {Gori},
  \citenamefont {Jung},\ and\ \citenamefont {Wang}}]{Gori:2013ala}%
  \BibitemOpen
  \bibfield  {author} {\bibinfo {author} {\bibfnamefont {S.}~\bibnamefont
  {Gori}}, \bibinfo {author} {\bibfnamefont {S.}~\bibnamefont {Jung}}, \ and\
  \bibinfo {author} {\bibfnamefont {L.-T.}\ \bibnamefont {Wang}},\ }\href@noop
  {} {\  (\bibinfo {year} {2013})},\ \Eprint {http://arxiv.org/abs/1307.5952}
  {arXiv:1307.5952 [hep-ph]} \BibitemShut {NoStop}%
%%CITATION = ARXIV:1307.5952;%%
\bibitem [{\citenamefont {Batell}\ \emph {et~al.}(2013)\citenamefont {Batell},
  \citenamefont {Jung},\ and\ \citenamefont {Wagner}}]{Batell:2013bka}%
  \BibitemOpen
  \bibfield  {author} {\bibinfo {author} {\bibfnamefont {B.}~\bibnamefont
  {Batell}}, \bibinfo {author} {\bibfnamefont {S.}~\bibnamefont {Jung}}, \ and\
  \bibinfo {author} {\bibfnamefont {C.~E.~M.}\ \bibnamefont {Wagner}},\ }\href
  {\doibase 10.1007/JHEP12(2013)075} {\  (\bibinfo {year} {2013}),\
  10.1007/JHEP12(2013)075},\ \Eprint {http://arxiv.org/abs/1309.2297}
  {arXiv:1309.2297 [hep-ph]} \BibitemShut {NoStop}%
%%CITATION = ARXIV:1309.2297;%%
\bibitem [{\citenamefont {Han}\ \emph {et~al.}(2013{\natexlab{b}})\citenamefont
  {Han}, \citenamefont {Padhi},\ and\ \citenamefont {Su}}]{Han:2013kza}%
  \BibitemOpen
  \bibfield  {author} {\bibinfo {author} {\bibfnamefont {T.}~\bibnamefont
  {Han}}, \bibinfo {author} {\bibfnamefont {S.}~\bibnamefont {Padhi}}, \ and\
  \bibinfo {author} {\bibfnamefont {S.}~\bibnamefont {Su}},\ }\href@noop {} {\
  (\bibinfo {year} {2013}{\natexlab{b}})},\ \Eprint
  {http://arxiv.org/abs/1309.5966} {arXiv:1309.5966 [hep-ph]} \BibitemShut
  {NoStop}%
%%CITATION = ARXIV:1309.5966;%%
\bibitem [{\citenamefont {Berggren}\ \emph
  {et~al.}(2013{\natexlab{b}})\citenamefont {Berggren}, \citenamefont {Han},
  \citenamefont {List}, \citenamefont {Padhi}, \citenamefont {Su} \emph
  {et~al.}}]{Berggren:2013bua}%
  \BibitemOpen
  \bibfield  {author} {\bibinfo {author} {\bibfnamefont {M.}~\bibnamefont
  {Berggren}}, \bibinfo {author} {\bibfnamefont {T.}~\bibnamefont {Han}},
  \bibinfo {author} {\bibfnamefont {J.}~\bibnamefont {List}}, \bibinfo {author}
  {\bibfnamefont {S.}~\bibnamefont {Padhi}}, \bibinfo {author} {\bibfnamefont
  {S.}~\bibnamefont {Su}},  \emph {et~al.},\ }\href@noop {} {\  (\bibinfo
  {year} {2013}{\natexlab{b}})},\ \Eprint {http://arxiv.org/abs/1309.7342}
  {arXiv:1309.7342 [hep-ph]} \BibitemShut {NoStop}%
%%CITATION = ARXIV:1309.7342;%%
\bibitem [{\citenamefont {Buckley}\ \emph {et~al.}(2013)\citenamefont
  {Buckley}, \citenamefont {Lykken}, \citenamefont {Rogan},\ and\ \citenamefont
  {Spiropulu}}]{Buckley:2013kua}%
  \BibitemOpen
  \bibfield  {author} {\bibinfo {author} {\bibfnamefont {M.~R.}\ \bibnamefont
  {Buckley}}, \bibinfo {author} {\bibfnamefont {J.~D.}\ \bibnamefont {Lykken}},
  \bibinfo {author} {\bibfnamefont {C.}~\bibnamefont {Rogan}}, \ and\ \bibinfo
  {author} {\bibfnamefont {M.}~\bibnamefont {Spiropulu}},\ }\href@noop {} {\
  (\bibinfo {year} {2013})},\ \Eprint {http://arxiv.org/abs/1310.4827}
  {arXiv:1310.4827 [hep-ph]} \BibitemShut {NoStop}%
%%CITATION = ARXIV:1310.4827;%%
\bibitem [{\citenamefont {Papaefstathiou}\ \emph {et~al.}(2014)\citenamefont
  {Papaefstathiou}, \citenamefont {Sakurai},\ and\ \citenamefont
  {Takeuchi}}]{Papaefstathiou:2014oja}%
  \BibitemOpen
  \bibfield  {author} {\bibinfo {author} {\bibfnamefont {A.}~\bibnamefont
  {Papaefstathiou}}, \bibinfo {author} {\bibfnamefont {K.}~\bibnamefont
  {Sakurai}}, \ and\ \bibinfo {author} {\bibfnamefont {M.}~\bibnamefont
  {Takeuchi}},\ }\href@noop {} {\  (\bibinfo {year} {2014})},\ \Eprint
  {http://arxiv.org/abs/1404.1077} {arXiv:1404.1077 [hep-ph]} \BibitemShut
  {NoStop}%
%%CITATION = ARXIV:1404.1077;%%
\bibitem [{\citenamefont {Chatrchyan}\ \emph
  {et~al.}(2013{\natexlab{k}})\citenamefont {Chatrchyan} \emph
  {et~al.}}]{Chatrchyan:2013mya}%
  \BibitemOpen
  \bibfield  {author} {\bibinfo {author} {\bibfnamefont {S.}~\bibnamefont
  {Chatrchyan}} \emph {et~al.} (\bibinfo {collaboration} {CMS Collaboration}),\
  }\href@noop {} {\  (\bibinfo {year} {2013}{\natexlab{k}})},\ \Eprint
  {http://arxiv.org/abs/1312.3310} {arXiv:1312.3310 [hep-ex]} \BibitemShut
  {NoStop}%
%%CITATION = ARXIV:1312.3310;%%
\bibitem [{\citenamefont {Howe}\ and\ \citenamefont
  {Saraswat}(2012)}]{Howe:2012xe}%
  \BibitemOpen
  \bibfield  {author} {\bibinfo {author} {\bibfnamefont {K.}~\bibnamefont
  {Howe}}\ and\ \bibinfo {author} {\bibfnamefont {P.}~\bibnamefont
  {Saraswat}},\ }\href {\doibase 10.1007/JHEP10(2012)065} {\bibfield  {journal}
  {\bibinfo  {journal} {JHEP}\ }\textbf {\bibinfo {volume} {1210}},\ \bibinfo
  {pages} {065} (\bibinfo {year} {2012})},\ \Eprint
  {http://arxiv.org/abs/1208.1542} {arXiv:1208.1542 [hep-ph]} \BibitemShut
  {NoStop}%
%%CITATION = ARXIV:1208.1542;%%
\bibitem [{\citenamefont {Kumar}\ \emph {et~al.}(2012)\citenamefont {Kumar},
  \citenamefont {Vega-Morales},\ and\ \citenamefont {Yu}}]{Kumar:2012ww}%
  \BibitemOpen
  \bibfield  {author} {\bibinfo {author} {\bibfnamefont {K.}~\bibnamefont
  {Kumar}}, \bibinfo {author} {\bibfnamefont {R.}~\bibnamefont {Vega-Morales}},
  \ and\ \bibinfo {author} {\bibfnamefont {F.}~\bibnamefont {Yu}},\ }\href
  {\doibase 10.1103/PhysRevD.86.113002} {\bibfield  {journal} {\bibinfo
  {journal} {Phys.Rev.}\ }\textbf {\bibinfo {volume} {D86}},\ \bibinfo {pages}
  {113002} (\bibinfo {year} {2012})},\ \Eprint {http://arxiv.org/abs/1205.4244}
  {arXiv:1205.4244 [hep-ph]} \BibitemShut {NoStop}%
%%CITATION = ARXIV:1205.4244;%%
\bibitem [{\citenamefont {Dittmaier}\ \emph {et~al.}(2011)\citenamefont
  {Dittmaier} \emph {et~al.}}]{Dittmaier:2011ti}%
  \BibitemOpen
  \bibfield  {author} {\bibinfo {author} {\bibfnamefont {S.}~\bibnamefont
  {Dittmaier}} \emph {et~al.} (\bibinfo {collaboration} {LHC Higgs Cross
  Section Working Group}),\ }\href {\doibase 10.5170/CERN-2011-002} {\
  (\bibinfo {year} {2011}),\ 10.5170/CERN-2011-002},\ \Eprint
  {http://arxiv.org/abs/1101.0593} {arXiv:1101.0593 [hep-ph]} \BibitemShut
  {NoStop}%
%%CITATION = ARXIV:1101.0593;%%
\bibitem [{\citenamefont {Dittmaier}\ \emph {et~al.}(2012)\citenamefont
  {Dittmaier}, \citenamefont {Dittmaier}, \citenamefont {Mariotti},
  \citenamefont {Passarino}, \citenamefont {Tanaka} \emph
  {et~al.}}]{Dittmaier:2012vm}%
  \BibitemOpen
  \bibfield  {author} {\bibinfo {author} {\bibfnamefont {S.}~\bibnamefont
  {Dittmaier}}, \bibinfo {author} {\bibfnamefont {S.}~\bibnamefont
  {Dittmaier}}, \bibinfo {author} {\bibfnamefont {C.}~\bibnamefont {Mariotti}},
  \bibinfo {author} {\bibfnamefont {G.}~\bibnamefont {Passarino}}, \bibinfo
  {author} {\bibfnamefont {R.}~\bibnamefont {Tanaka}},  \emph {et~al.},\ }\href
  {\doibase 10.5170/CERN-2012-002} {\  (\bibinfo {year} {2012}),\
  10.5170/CERN-2012-002},\ \Eprint {http://arxiv.org/abs/1201.3084}
  {arXiv:1201.3084 [hep-ph]} \BibitemShut {NoStop}%
%%CITATION = ARXIV:1201.3084;%%
\bibitem [{\citenamefont {Heinemeyer}\ \emph {et~al.}(2013)\citenamefont
  {Heinemeyer} \emph {et~al.}}]{Heinemeyer:2013tqa}%
  \BibitemOpen
  \bibfield  {author} {\bibinfo {author} {\bibfnamefont {S.}~\bibnamefont
  {Heinemeyer}} \emph {et~al.} (\bibinfo {collaboration} {LHC Higgs Cross
  Section Working Group}),\ }\href {\doibase 10.5170/CERN-2013-004} {\
  (\bibinfo {year} {2013}),\ 10.5170/CERN-2013-004},\ \Eprint
  {http://arxiv.org/abs/1307.1347} {arXiv:1307.1347 [hep-ph]} \BibitemShut
  {NoStop}%
%%CITATION = ARXIV:1307.1347;%%
\bibitem [{\citenamefont {Jaiswal}\ \emph {et~al.}(2013)\citenamefont
  {Jaiswal}, \citenamefont {Kopp},\ and\ \citenamefont
  {Okui}}]{Jaiswal:2013xra}%
  \BibitemOpen
  \bibfield  {author} {\bibinfo {author} {\bibfnamefont {P.}~\bibnamefont
  {Jaiswal}}, \bibinfo {author} {\bibfnamefont {K.}~\bibnamefont {Kopp}}, \
  and\ \bibinfo {author} {\bibfnamefont {T.}~\bibnamefont {Okui}},\ }\href
  {\doibase 10.1103/PhysRevD.87.115017} {\bibfield  {journal} {\bibinfo
  {journal} {Phys.Rev.}\ }\textbf {\bibinfo {volume} {D87}},\ \bibinfo {pages}
  {115017} (\bibinfo {year} {2013})},\ \Eprint {http://arxiv.org/abs/1303.1181}
  {arXiv:1303.1181 [hep-ph]} \BibitemShut {NoStop}%
%%CITATION = ARXIV:1303.1181;%%
\bibitem [{\citenamefont {Chatrchyan}\ \emph
  {et~al.}(2013{\natexlab{l}})\citenamefont {Chatrchyan} \emph
  {et~al.}}]{CMS:2013yea}%
  \BibitemOpen
  \bibfield  {author} {\bibinfo {author} {\bibfnamefont {S.}~\bibnamefont
  {Chatrchyan}} \emph {et~al.} (\bibinfo {collaboration} {CMS Collaboration}),\
  }\href@noop {} {\  (\bibinfo {year} {2013}{\natexlab{l}})},\ \bibinfo {note}
  {{C}MS-PAS-HIG-13-022}\BibitemShut {NoStop}%
%%CITATION = CMS-PAS-HIG-13-022 ETC.;%%
\bibitem [{\citenamefont {Chatrchyan}\ \emph
  {et~al.}(2013{\natexlab{m}})\citenamefont {Chatrchyan} \emph
  {et~al.}}]{CMS:2013jda}%
  \BibitemOpen
  \bibfield  {author} {\bibinfo {author} {\bibfnamefont {S.}~\bibnamefont
  {Chatrchyan}} \emph {et~al.} (\bibinfo {collaboration} {CMS Collaboration}),\
  }\href@noop {} {\  (\bibinfo {year} {2013}{\natexlab{m}})},\ \bibinfo {note}
  {{C}MS-PAS-HIG-13-011}\BibitemShut {NoStop}%
%%CITATION = CMS-PAS-HIG-13-011 ETC.;%%
\bibitem [{\citenamefont {Chatrchyan}\ \emph
  {et~al.}(2013{\natexlab{n}})\citenamefont {Chatrchyan} \emph
  {et~al.}}]{CMS:ckv}%
  \BibitemOpen
  \bibfield  {author} {\bibinfo {author} {\bibfnamefont {S.}~\bibnamefont
  {Chatrchyan}} \emph {et~al.} (\bibinfo {collaboration} {CMS Collaboration}),\
  }\href@noop {} {\  (\bibinfo {year} {2013}{\natexlab{n}})},\ \bibinfo {note}
  {{C}MS-PAS-HIG-12-053}\BibitemShut {NoStop}%
%%CITATION = CMS-PAS-HIG-12-053 ETC.;%%
\bibitem [{\citenamefont {Chatrchyan}\ \emph
  {et~al.}(2013{\natexlab{o}})\citenamefont {Chatrchyan} \emph
  {et~al.}}]{CMS:zwa}%
  \BibitemOpen
  \bibfield  {author} {\bibinfo {author} {\bibfnamefont {S.}~\bibnamefont
  {Chatrchyan}} \emph {et~al.} (\bibinfo {collaboration} {CMS Collaboration}),\
  }\href@noop {} {\  (\bibinfo {year} {2013}{\natexlab{o}})},\ \bibinfo {note}
  {{C}MS-PAS-HIG-13-009}\BibitemShut {NoStop}%
%%CITATION = CMS-PAS-HIG-13-009 ETC.;%%
\bibitem [{\citenamefont {Chatrchyan}\ \emph
  {et~al.}(2013{\natexlab{p}})\citenamefont {Chatrchyan} \emph
  {et~al.}}]{Chatrchyan:2013zna}%
  \BibitemOpen
  \bibfield  {author} {\bibinfo {author} {\bibfnamefont {S.}~\bibnamefont
  {Chatrchyan}} \emph {et~al.} (\bibinfo {collaboration} {CMS Collaboration}),\
  }\href@noop {} {\  (\bibinfo {year} {2013}{\natexlab{p}})},\ \Eprint
  {http://arxiv.org/abs/1310.3687} {arXiv:1310.3687 [hep-ex]} \BibitemShut
  {NoStop}%
%%CITATION = ARXIV:1310.3687;%%
\bibitem [{\citenamefont {Chatrchyan}\ \emph
  {et~al.}(2013{\natexlab{q}})\citenamefont {Chatrchyan} \emph
  {et~al.}}]{CMS:2013xda}%
  \BibitemOpen
  \bibfield  {author} {\bibinfo {author} {\bibfnamefont {S.}~\bibnamefont
  {Chatrchyan}} \emph {et~al.} (\bibinfo {collaboration} {CMS Collaboration}),\
  }\href@noop {} {\  (\bibinfo {year} {2013}{\natexlab{q}})},\ \bibinfo {note}
  {{C}MS-PAS-HIG-13-017}\BibitemShut {NoStop}%
%%CITATION = CMS-PAS-HIG-13-017 ETC.;%%
\bibitem [{\citenamefont {Aad}\ \emph {et~al.}(2013{\natexlab{m}})\citenamefont
  {Aad} \emph {et~al.}}]{TheATLAScollaboration:2013hia}%
  \BibitemOpen
  \bibfield  {author} {\bibinfo {author} {\bibfnamefont {G.}~\bibnamefont
  {Aad}} \emph {et~al.} (\bibinfo {collaboration} {ATLAS Collaboration}),\
  }\href@noop {} {\  (\bibinfo {year} {2013}{\natexlab{m}})},\ \bibinfo {note}
  {{A}TLAS-CONF-2013-075}\BibitemShut {NoStop}%
%%CITATION = ATLAS-CONF-2013-075 ETC.;%%
\bibitem [{\citenamefont {Aad}\ \emph {et~al.}(2013{\natexlab{n}})\citenamefont
  {Aad} \emph {et~al.}}]{TheATLAScollaboration:2013lia}%
  \BibitemOpen
  \bibfield  {author} {\bibinfo {author} {\bibfnamefont {G.}~\bibnamefont
  {Aad}} \emph {et~al.} (\bibinfo {collaboration} {ATLAS Collaboration}),\
  }\href@noop {} {\  (\bibinfo {year} {2013}{\natexlab{n}})},\ \bibinfo {note}
  {{A}TLAS-CONF-2013-079}\BibitemShut {NoStop}%
%%CITATION = ATLAS-CONF-2013-079 ETC.;%%
\bibitem [{\citenamefont {Chatrchyan}\ \emph
  {et~al.}(2013{\natexlab{r}})\citenamefont {Chatrchyan} \emph
  {et~al.}}]{CMS:2013sea}%
  \BibitemOpen
  \bibfield  {author} {\bibinfo {author} {\bibfnamefont {S.}~\bibnamefont
  {Chatrchyan}} \emph {et~al.} (\bibinfo {collaboration} {CMS Collaboration}),\
  }\href@noop {} {\  (\bibinfo {year} {2013}{\natexlab{r}})},\ \bibinfo {note}
  {{C}MS-PAS-HIG-13-019}\BibitemShut {NoStop}%
%%CITATION = CMS-PAS-HIG-13-019 ETC.;%%
\bibitem [{\citenamefont {Chatrchyan}\ \emph
  {et~al.}(2013{\natexlab{s}})\citenamefont {Chatrchyan} \emph
  {et~al.}}]{CMS:2013fda}%
  \BibitemOpen
  \bibfield  {author} {\bibinfo {author} {\bibfnamefont {S.}~\bibnamefont
  {Chatrchyan}} \emph {et~al.} (\bibinfo {collaboration} {CMS Collaboration}),\
  }\href@noop {} {\  (\bibinfo {year} {2013}{\natexlab{s}})},\ \bibinfo {note}
  {{C}MS-PAS-HIG-13-015}\BibitemShut {NoStop}%
%%CITATION = CMS-PAS-HIG-13-015 ETC.;%%
\bibitem [{\citenamefont {Chatrchyan}\ \emph
  {et~al.}(2013{\natexlab{t}})\citenamefont {Chatrchyan} \emph
  {et~al.}}]{CMS:2013tfa}%
  \BibitemOpen
  \bibfield  {author} {\bibinfo {author} {\bibfnamefont {S.}~\bibnamefont
  {Chatrchyan}} \emph {et~al.} (\bibinfo {collaboration} {CMS Collaboration}),\
  }\href@noop {} {\  (\bibinfo {year} {2013}{\natexlab{t}})},\ \bibinfo {note}
  {{C}MS-PAS-HIG-13-020}\BibitemShut {NoStop}%
%%CITATION = CMS-PAS-HIG-13-020 ETC.;%%
\bibitem [{\citenamefont {Aad}\ \emph {et~al.}(2013{\natexlab{o}})\citenamefont
  {Aad} \emph {et~al.}}]{TheATLAScollaboration:2013mia}%
  \BibitemOpen
  \bibfield  {author} {\bibinfo {author} {\bibfnamefont {G.}~\bibnamefont
  {Aad}} \emph {et~al.} (\bibinfo {collaboration} {ATLAS Collaboration}),\
  }\href@noop {} {\  (\bibinfo {year} {2013}{\natexlab{o}})},\ \bibinfo {note}
  {{A}TLAS-CONF-2013-080}\BibitemShut {NoStop}%
%%CITATION = ATLAS-CONF-2013-080 ETC.;%%
\bibitem [{\citenamefont {Boudjema}\ \emph {et~al.}(2013)\citenamefont
  {Boudjema}, \citenamefont {Cacciapaglia}, \citenamefont {Cranmer},
  \citenamefont {Dissertori}, \citenamefont {Deandrea} \emph
  {et~al.}}]{Boudjema:2013qla}%
  \BibitemOpen
  \bibfield  {author} {\bibinfo {author} {\bibfnamefont {F.}~\bibnamefont
  {Boudjema}}, \bibinfo {author} {\bibfnamefont {G.}~\bibnamefont
  {Cacciapaglia}}, \bibinfo {author} {\bibfnamefont {K.}~\bibnamefont
  {Cranmer}}, \bibinfo {author} {\bibfnamefont {G.}~\bibnamefont {Dissertori}},
  \bibinfo {author} {\bibfnamefont {A.}~\bibnamefont {Deandrea}},  \emph
  {et~al.},\ }\href@noop {} {\  (\bibinfo {year} {2013})},\ \Eprint
  {http://arxiv.org/abs/1307.5865} {arXiv:1307.5865 [hep-ph]} \BibitemShut
  {NoStop}%
%%CITATION = ARXIV:1307.5865;%%
\bibitem [{\citenamefont {Aad}\ \emph {et~al.}(2013{\natexlab{p}})\citenamefont
  {Aad} \emph {et~al.}}]{TheATLAScollaboration:2013eia}%
  \BibitemOpen
  \bibfield  {author} {\bibinfo {author} {\bibfnamefont {G.}~\bibnamefont
  {Aad}} \emph {et~al.},\ }\href@noop {} {\  (\bibinfo {year}
  {2013}{\natexlab{p}})},\ \bibinfo {note} {{A}TLAS-CONF-2013-072}\BibitemShut
  {NoStop}%
%%CITATION = ATLAS-CONF-2013-072 ETC.;%%
\bibitem [{\citenamefont {Allanach}(2002)}]{Allanach:2001kg}%
  \BibitemOpen
  \bibfield  {author} {\bibinfo {author} {\bibfnamefont {B.}~\bibnamefont
  {Allanach}},\ }\href {\doibase 10.1016/S0010-4655(01)00460-X} {\bibfield
  {journal} {\bibinfo  {journal} {Comput.Phys.Commun.}\ }\textbf {\bibinfo
  {volume} {143}},\ \bibinfo {pages} {305} (\bibinfo {year} {2002})},\ \Eprint
  {http://arxiv.org/abs/hep-ph/0104145} {arXiv:hep-ph/0104145 [hep-ph]}
  \BibitemShut {NoStop}%
%%CITATION = HEP-PH/0104145;%%
\bibitem [{\citenamefont {Djouadi}\ \emph
  {et~al.}(2007{\natexlab{a}})\citenamefont {Djouadi}, \citenamefont
  {Muhlleitner},\ and\ \citenamefont {Spira}}]{Djouadi:2006bz}%
  \BibitemOpen
  \bibfield  {author} {\bibinfo {author} {\bibfnamefont {A.}~\bibnamefont
  {Djouadi}}, \bibinfo {author} {\bibfnamefont {M.}~\bibnamefont
  {Muhlleitner}}, \ and\ \bibinfo {author} {\bibfnamefont {M.}~\bibnamefont
  {Spira}},\ }\href@noop {} {\bibfield  {journal} {\bibinfo  {journal} {Acta
  Phys.Polon.}\ }\textbf {\bibinfo {volume} {B38}},\ \bibinfo {pages} {635}
  (\bibinfo {year} {2007}{\natexlab{a}})},\ \Eprint
  {http://arxiv.org/abs/hep-ph/0609292} {arXiv:hep-ph/0609292 [hep-ph]}
  \BibitemShut {NoStop}%
%%CITATION = HEP-PH/0609292;%%
\bibitem [{\citenamefont {Djouadi}\ \emph
  {et~al.}(2007{\natexlab{b}})\citenamefont {Djouadi}, \citenamefont {Kneur},\
  and\ \citenamefont {Moultaka}}]{Djouadi:2002ze}%
  \BibitemOpen
  \bibfield  {author} {\bibinfo {author} {\bibfnamefont {A.}~\bibnamefont
  {Djouadi}}, \bibinfo {author} {\bibfnamefont {J.-L.}\ \bibnamefont {Kneur}},
  \ and\ \bibinfo {author} {\bibfnamefont {G.}~\bibnamefont {Moultaka}},\
  }\href {\doibase 10.1016/j.cpc.2006.11.009} {\bibfield  {journal} {\bibinfo
  {journal} {Comput.Phys.Commun.}\ }\textbf {\bibinfo {volume} {176}},\
  \bibinfo {pages} {426} (\bibinfo {year} {2007}{\natexlab{b}})},\ \Eprint
  {http://arxiv.org/abs/hep-ph/0211331} {arXiv:hep-ph/0211331 [hep-ph]}
  \BibitemShut {NoStop}%
%%CITATION = HEP-PH/0211331;%%
\bibitem [{\citenamefont {Djouadi}\ \emph {et~al.}(1998)\citenamefont
  {Djouadi}, \citenamefont {Kalinowski},\ and\ \citenamefont
  {Spira}}]{Djouadi:1997yw}%
  \BibitemOpen
  \bibfield  {author} {\bibinfo {author} {\bibfnamefont {A.}~\bibnamefont
  {Djouadi}}, \bibinfo {author} {\bibfnamefont {J.}~\bibnamefont {Kalinowski}},
  \ and\ \bibinfo {author} {\bibfnamefont {M.}~\bibnamefont {Spira}},\ }\href
  {\doibase 10.1016/S0010-4655(97)00123-9} {\bibfield  {journal} {\bibinfo
  {journal} {Comput.Phys.Commun.}\ }\textbf {\bibinfo {volume} {108}},\
  \bibinfo {pages} {56} (\bibinfo {year} {1998})},\ \Eprint
  {http://arxiv.org/abs/hep-ph/9704448} {arXiv:hep-ph/9704448 [hep-ph]}
  \BibitemShut {NoStop}%
%%CITATION = HEP-PH/9704448;%%
\bibitem [{\citenamefont {Muhlleitner}\ \emph {et~al.}(2005)\citenamefont
  {Muhlleitner}, \citenamefont {Djouadi},\ and\ \citenamefont
  {Mambrini}}]{Muhlleitner:2003vg}%
  \BibitemOpen
  \bibfield  {author} {\bibinfo {author} {\bibfnamefont {M.}~\bibnamefont
  {Muhlleitner}}, \bibinfo {author} {\bibfnamefont {A.}~\bibnamefont
  {Djouadi}}, \ and\ \bibinfo {author} {\bibfnamefont {Y.}~\bibnamefont
  {Mambrini}},\ }\href {\doibase 10.1016/j.cpc.2005.01.012} {\bibfield
  {journal} {\bibinfo  {journal} {Comput.Phys.Commun.}\ }\textbf {\bibinfo
  {volume} {168}},\ \bibinfo {pages} {46} (\bibinfo {year} {2005})},\ \Eprint
  {http://arxiv.org/abs/hep-ph/0311167} {arXiv:hep-ph/0311167 [hep-ph]}
  \BibitemShut {NoStop}%
%%CITATION = HEP-PH/0311167;%%
\bibitem [{\citenamefont {Beenakker}\ \emph {et~al.}(1999)\citenamefont
  {Beenakker}, \citenamefont {Klasen}, \citenamefont {Kramer}, \citenamefont
  {Plehn}, \citenamefont {Spira} \emph {et~al.}}]{Beenakker:1999xh}%
  \BibitemOpen
  \bibfield  {author} {\bibinfo {author} {\bibfnamefont {W.}~\bibnamefont
  {Beenakker}}, \bibinfo {author} {\bibfnamefont {M.}~\bibnamefont {Klasen}},
  \bibinfo {author} {\bibfnamefont {M.}~\bibnamefont {Kramer}}, \bibinfo
  {author} {\bibfnamefont {T.}~\bibnamefont {Plehn}}, \bibinfo {author}
  {\bibfnamefont {M.}~\bibnamefont {Spira}},  \emph {et~al.},\ }\href {\doibase
  10.1103/PhysRevLett.100.029901, 10.1103/PhysRevLett.83.3780} {\bibfield
  {journal} {\bibinfo  {journal} {Phys.Rev.Lett.}\ }\textbf {\bibinfo {volume}
  {83}},\ \bibinfo {pages} {3780} (\bibinfo {year} {1999})},\ \Eprint
  {http://arxiv.org/abs/hep-ph/9906298} {arXiv:hep-ph/9906298 [hep-ph]}
  \BibitemShut {NoStop}%
%%CITATION = HEP-PH/9906298;%%
\bibitem [{\citenamefont {Belanger}\ \emph {et~al.}(2014)\citenamefont
  {Belanger}, \citenamefont {Boudjema}, \citenamefont {Pukhov},\ and\
  \citenamefont {Semenov}}]{Belanger:2013oya}%
  \BibitemOpen
  \bibfield  {author} {\bibinfo {author} {\bibfnamefont {G.}~\bibnamefont
  {Belanger}}, \bibinfo {author} {\bibfnamefont {F.}~\bibnamefont {Boudjema}},
  \bibinfo {author} {\bibfnamefont {A.}~\bibnamefont {Pukhov}}, \ and\ \bibinfo
  {author} {\bibfnamefont {A.}~\bibnamefont {Semenov}},\ }\href {\doibase
  10.1016/j.cpc.2013.10.016} {\bibfield  {journal} {\bibinfo  {journal}
  {Comput.Phys.Commun.}\ }\textbf {\bibinfo {volume} {185}},\ \bibinfo {pages}
  {960} (\bibinfo {year} {2014})},\ \Eprint {http://arxiv.org/abs/1305.0237}
  {arXiv:1305.0237 [hep-ph]} \BibitemShut {NoStop}%
%%CITATION = ARXIV:1305.0237;%%
\bibitem [{\citenamefont {Beringer}\ \emph {et~al.}(2012)\citenamefont
  {Beringer} \emph {et~al.}}]{Beringer:1900zz}%
  \BibitemOpen
  \bibfield  {author} {\bibinfo {author} {\bibfnamefont {J.}~\bibnamefont
  {Beringer}} \emph {et~al.} (\bibinfo {collaboration} {Particle Data Group}),\
  }\href {\doibase 10.1103/PhysRevD.86.010001} {\bibfield  {journal} {\bibinfo
  {journal} {Phys.Rev.}\ }\textbf {\bibinfo {volume} {D86}},\ \bibinfo {pages}
  {010001} (\bibinfo {year} {2012})}\BibitemShut {NoStop}%
%%CITATION = PHRVA,D86,010001;%%
\bibitem [{\citenamefont {Akerib}\ \emph {et~al.}(2013)\citenamefont {Akerib}
  \emph {et~al.}}]{Akerib:2013tjd}%
  \BibitemOpen
  \bibfield  {author} {\bibinfo {author} {\bibfnamefont {D.}~\bibnamefont
  {Akerib}} \emph {et~al.} (\bibinfo {collaboration} {LUX Collaboration}),\
  }\href@noop {} {\  (\bibinfo {year} {2013})},\ \Eprint
  {http://arxiv.org/abs/1310.8214} {arXiv:1310.8214 [astro-ph.CO]} \BibitemShut
  {NoStop}%
%%CITATION = ARXIV:1310.8214;%%
\bibitem [{\citenamefont {Alves}\ \emph {et~al.}(2012)\citenamefont {Alves}
  \emph {et~al.}}]{Alves:2011wf}%
  \BibitemOpen
  \bibfield  {author} {\bibinfo {author} {\bibfnamefont {D.}~\bibnamefont
  {Alves}} \emph {et~al.} (\bibinfo {collaboration} {LHC New Physics Working
  Group}),\ }\href {\doibase 10.1088/0954-3899/39/10/105005} {\bibfield
  {journal} {\bibinfo  {journal} {J.Phys.}\ }\textbf {\bibinfo {volume}
  {G39}},\ \bibinfo {pages} {105005} (\bibinfo {year} {2012})},\ \Eprint
  {http://arxiv.org/abs/1105.2838} {arXiv:1105.2838 [hep-ph]} \BibitemShut
  {NoStop}%
%%CITATION = ARXIV:1105.2838;%%
\bibitem [{\citenamefont {Aaltonen}\ \emph {et~al.}(2013)\citenamefont
  {Aaltonen} \emph {et~al.}}]{Aaltonen:2013vca}%
  \BibitemOpen
  \bibfield  {author} {\bibinfo {author} {\bibfnamefont {T.~A.}\ \bibnamefont
  {Aaltonen}} \emph {et~al.} (\bibinfo {collaboration} {CDF Collaboration}),\
  }\href@noop {} {\bibfield  {journal} {\bibinfo  {journal} {Phys.Rev.Lett.}\ }
  (\bibinfo {year} {2013})},\ \Eprint {http://arxiv.org/abs/1309.7509}
  {arXiv:1309.7509 [hep-ex]} \BibitemShut {NoStop}%
%%CITATION = ARXIV:1309.7509;%%
\bibitem [{\citenamefont {Aad}\ \emph {et~al.}(2013{\natexlab{q}})\citenamefont
  {Aad} \emph {et~al.}}]{Aad:2012xsa}%
  \BibitemOpen
  \bibfield  {author} {\bibinfo {author} {\bibfnamefont {G.}~\bibnamefont
  {Aad}} \emph {et~al.} (\bibinfo {collaboration} {ATLAS Collaboration}),\
  }\href {\doibase 10.1103/PhysRevD.87.052002} {\bibfield  {journal} {\bibinfo
  {journal} {Phys.Rev.}\ }\textbf {\bibinfo {volume} {D87}},\ \bibinfo {pages}
  {052002} (\bibinfo {year} {2013}{\natexlab{q}})},\ \Eprint
  {http://arxiv.org/abs/1211.6312} {arXiv:1211.6312 [hep-ex]} \BibitemShut
  {NoStop}%
%%CITATION = ARXIV:1211.6312;%%
\bibitem [{\citenamefont {Aad}\ \emph {et~al.}(2014{\natexlab{b}})\citenamefont
  {Aad} \emph {et~al.}}]{Aad:2014nua}%
  \BibitemOpen
  \bibfield  {author} {\bibinfo {author} {\bibfnamefont {G.}~\bibnamefont
  {Aad}} \emph {et~al.} (\bibinfo {collaboration} {ATLAS Collaboration}),\
  }\href@noop {} {\  (\bibinfo {year} {2014}{\natexlab{b}})},\ \Eprint
  {http://arxiv.org/abs/1402.7029} {arXiv:1402.7029 [hep-ex]} \BibitemShut
  {NoStop}%
%%CITATION = ARXIV:1402.7029;%%
\bibitem [{\citenamefont {Aad}\ \emph {et~al.}(2013{\natexlab{r}})\citenamefont
  {Aad} \emph {et~al.}}]{TheATLAScollaboration:2013hha}%
  \BibitemOpen
  \bibfield  {author} {\bibinfo {author} {\bibfnamefont {G.}~\bibnamefont
  {Aad}} \emph {et~al.} (\bibinfo {collaboration} {ATLAS Collaboration}),\
  }\href@noop {} {\  (\bibinfo {year} {2013}{\natexlab{r}})},\ \bibinfo {note}
  {{A}TLAS-CONF-2013-049}\BibitemShut {NoStop}%
%%CITATION = ATLAS-CONF-2013-049 ETC.;%%
\bibitem [{\citenamefont {Chatrchyan}\ \emph
  {et~al.}(2013{\natexlab{u}})\citenamefont {Chatrchyan} \emph
  {et~al.}}]{CMS:2013dea}%
  \BibitemOpen
  \bibfield  {author} {\bibinfo {author} {\bibfnamefont {S.}~\bibnamefont
  {Chatrchyan}} \emph {et~al.} (\bibinfo {collaboration} {CMS Collaboration}),\
  }\href@noop {} {\  (\bibinfo {year} {2013}{\natexlab{u}})},\ \bibinfo {note}
  {{C}MS-PAS-SUS-13-006}\BibitemShut {NoStop}%
%%CITATION = CMS-PAS-SUS-13-006 ETC.;%%
\bibitem [{\citenamefont {Aad}\ \emph {et~al.}(2013{\natexlab{s}})\citenamefont
  {Aad} \emph {et~al.}}]{TheATLAScollaboration:2013zia}%
  \BibitemOpen
  \bibfield  {author} {\bibinfo {author} {\bibfnamefont {G.}~\bibnamefont
  {Aad}} \emph {et~al.} (\bibinfo {collaboration} {ATLAS Collaboration}),\
  }\href@noop {} {\  (\bibinfo {year} {2013}{\natexlab{s}})},\ \bibinfo {note}
  {{A}TLAS-CONF-2013-093}\BibitemShut {NoStop}%
%%CITATION = ATLAS-CONF-2013-093 ETC.;%%
\bibitem [{\citenamefont {Chatrchyan}\ \emph
  {et~al.}(2013{\natexlab{v}})\citenamefont {Chatrchyan} \emph
  {et~al.}}]{CMS:2013afa}%
  \BibitemOpen
  \bibfield  {author} {\bibinfo {author} {\bibfnamefont {S.}~\bibnamefont
  {Chatrchyan}} \emph {et~al.} (\bibinfo {collaboration} {CMS Collaboration}),\
  }\href@noop {} {\  (\bibinfo {year} {2013}{\natexlab{v}})},\ \bibinfo {note}
  {{C}MS-PAS-SUS-13-017}\BibitemShut {NoStop}%
%%CITATION = CMS-PAS-SUS-13-017 ETC.;%%
\bibitem [{\citenamefont {Alwall}\ \emph {et~al.}(2011)\citenamefont {Alwall},
  \citenamefont {Herquet}, \citenamefont {Maltoni}, \citenamefont {Mattelaer},\
  and\ \citenamefont {Stelzer}}]{Alwall:2011uj}%
  \BibitemOpen
  \bibfield  {author} {\bibinfo {author} {\bibfnamefont {J.}~\bibnamefont
  {Alwall}}, \bibinfo {author} {\bibfnamefont {M.}~\bibnamefont {Herquet}},
  \bibinfo {author} {\bibfnamefont {F.}~\bibnamefont {Maltoni}}, \bibinfo
  {author} {\bibfnamefont {O.}~\bibnamefont {Mattelaer}}, \ and\ \bibinfo
  {author} {\bibfnamefont {T.}~\bibnamefont {Stelzer}},\ }\href {\doibase
  10.1007/JHEP06(2011)128} {\bibfield  {journal} {\bibinfo  {journal} {JHEP}\
  }\textbf {\bibinfo {volume} {1106}},\ \bibinfo {pages} {128} (\bibinfo {year}
  {2011})},\ \Eprint {http://arxiv.org/abs/1106.0522} {arXiv:1106.0522
  [hep-ph]} \BibitemShut {NoStop}%
%%CITATION = ARXIV:1106.0522;%%
\bibitem [{\citenamefont {Pumplin}\ \emph {et~al.}(2002)\citenamefont
  {Pumplin}, \citenamefont {Stump}, \citenamefont {Huston}, \citenamefont
  {Lai}, \citenamefont {Nadolsky} \emph {et~al.}}]{Pumplin:2002vw}%
  \BibitemOpen
  \bibfield  {author} {\bibinfo {author} {\bibfnamefont {J.}~\bibnamefont
  {Pumplin}}, \bibinfo {author} {\bibfnamefont {D.}~\bibnamefont {Stump}},
  \bibinfo {author} {\bibfnamefont {J.}~\bibnamefont {Huston}}, \bibinfo
  {author} {\bibfnamefont {H.}~\bibnamefont {Lai}}, \bibinfo {author}
  {\bibfnamefont {P.~M.}\ \bibnamefont {Nadolsky}},  \emph {et~al.},\ }\href
  {\doibase 10.1088/1126-6708/2002/07/012} {\bibfield  {journal} {\bibinfo
  {journal} {JHEP}\ }\textbf {\bibinfo {volume} {0207}},\ \bibinfo {pages}
  {012} (\bibinfo {year} {2002})},\ \Eprint
  {http://arxiv.org/abs/hep-ph/0201195} {arXiv:hep-ph/0201195 [hep-ph]}
  \BibitemShut {NoStop}%
%%CITATION = HEP-PH/0201195;%%
\bibitem [{\citenamefont {Sjostrand}\ \emph {et~al.}(2006)\citenamefont
  {Sjostrand}, \citenamefont {Mrenna},\ and\ \citenamefont
  {Skands}}]{Sjostrand:2006za}%
  \BibitemOpen
  \bibfield  {author} {\bibinfo {author} {\bibfnamefont {T.}~\bibnamefont
  {Sjostrand}}, \bibinfo {author} {\bibfnamefont {S.}~\bibnamefont {Mrenna}}, \
  and\ \bibinfo {author} {\bibfnamefont {P.~Z.}\ \bibnamefont {Skands}},\
  }\href {\doibase 10.1088/1126-6708/2006/05/026} {\bibfield  {journal}
  {\bibinfo  {journal} {JHEP}\ }\textbf {\bibinfo {volume} {0605}},\ \bibinfo
  {pages} {026} (\bibinfo {year} {2006})},\ \Eprint
  {http://arxiv.org/abs/hep-ph/0603175} {arXiv:hep-ph/0603175 [hep-ph]}
  \BibitemShut {NoStop}%
%%CITATION = HEP-PH/0603175;%%
\bibitem [{\citenamefont {Anastasiou}\ \emph {et~al.}(2014)\citenamefont
  {Anastasiou}, \citenamefont {Duhr}, \citenamefont {Dulat}, \citenamefont
  {Furlan}, \citenamefont {Gehrmann} \emph {et~al.}}]{Anastasiou:2014vaa}%
  \BibitemOpen
  \bibfield  {author} {\bibinfo {author} {\bibfnamefont {C.}~\bibnamefont
  {Anastasiou}}, \bibinfo {author} {\bibfnamefont {C.}~\bibnamefont {Duhr}},
  \bibinfo {author} {\bibfnamefont {F.}~\bibnamefont {Dulat}}, \bibinfo
  {author} {\bibfnamefont {E.}~\bibnamefont {Furlan}}, \bibinfo {author}
  {\bibfnamefont {T.}~\bibnamefont {Gehrmann}},  \emph {et~al.},\ }\href@noop
  {} {\  (\bibinfo {year} {2014})},\ \Eprint {http://arxiv.org/abs/1403.4616}
  {arXiv:1403.4616 [hep-ph]} \BibitemShut {NoStop}%
%%CITATION = ARXIV:1403.4616;%%
\bibitem [{\citenamefont {Djouadi}(2013)}]{Djouadi:2012rh}%
  \BibitemOpen
  \bibfield  {author} {\bibinfo {author} {\bibfnamefont {A.}~\bibnamefont
  {Djouadi}},\ }\href {\doibase 10.1140/epjc/s10052-013-2498-3} {\bibfield
  {journal} {\bibinfo  {journal} {Eur.Phys.J.}\ }\textbf {\bibinfo {volume}
  {C73}},\ \bibinfo {pages} {2498} (\bibinfo {year} {2013})},\ \Eprint
  {http://arxiv.org/abs/1208.3436} {arXiv:1208.3436 [hep-ph]} \BibitemShut
  {NoStop}%
%%CITATION = ARXIV:1208.3436;%%
\end{thebibliography}%

%\begin{thebibliography}{99}
%\end{thebibliography} 
\end{document}